\documentclass[12pt,a4paper]{article}
\usepackage{multicol}
\usepackage[english]{babel}
\usepackage[a4paper,top=2cm,bottom=2cm,left=2.5cm,right=2.5cm,marginparwidth=1.75cm]{geometry}

\usepackage{sectsty}
\allsectionsfont{\normalfont\normalsize\bfseries}

\usepackage{natbib}
\usepackage{amsmath}
\usepackage{graphicx}
\usepackage{hyperref}
\usepackage{color}
\usepackage{amsmath}
\usepackage{amssymb}
\usepackage{mathabx}

\usepackage{xcolor}
\definecolor{deeppink}{rgb}{1.0, 0.08, 0.58}
\hypersetup{
    colorlinks=true,
    linkcolor=deeppink,
    citecolor=deeppink,
    urlcolor=deeppink}

\usepackage{orcidlink}
\usepackage[title]{appendix}
\usepackage{mathrsfs}
\usepackage{amsfonts}
\usepackage{booktabs} 
\usepackage{caption}  
\usepackage{threeparttable} 
\usepackage{algorithm}
\usepackage{algorithmicx}
\usepackage{algpseudocode}
\usepackage{listings}
\usepackage{enumitem}
\usepackage{chngcntr}
\usepackage{booktabs}
\usepackage{lipsum}
\usepackage{subcaption}
\usepackage{authblk}
\usepackage[T1]{fontenc}  
\usepackage{csquotes}      
\usepackage{diagbox}

\usepackage{titlesec}

\titleformat{\section}
  {\normalfont\bfseries\normalsize}{\thesection}{1em}{}
  
\usepackage{float}   
\usepackage{caption} 


\usepackage{geometry}
\date{}

\title{Bulk Flow Motion Detection in the Local Universe with Pantheon$+$ Type Ia Supernovae}

\author[1]{Maria Lopes\orcidlink{0000-0001-9181-5675}}
\author[1]{Armando Bernui\orcidlink{0000-0003-3034-0762}}
\author[1]{Camila Franco\orcidlink{0000-0002-6320-425X}}
\author[1]{Felipe Avila\orcidlink{0000-0002-0562-2541}}
\affil[1]{\small Observatório Nacional, Rua General José Cristino 77, São Cristóvão, 20921-400 Rio de Janeiro,
RJ, Brasil}

\affil[ ]{\href{mailto:marialopes@on.br}{\textcolor{deeppink}{\texttt{marialopes@on.br}}}, \href{mailto:bernui@on.br}{\textcolor{deeppink}{\texttt{bernui@on.br}}}, \href{mailto:camilafranco@on.br}{\textcolor{deeppink}{\texttt{camilafranco@on.br}}}, \href{mailto:fsavila2@gmail.com}{\textcolor{deeppink}{\texttt{fsavila2@gmail.com}}}}

\begin{document}
\maketitle
\vspace{-2.5\baselineskip}

\begin{center}
(Received 19-Mar, 2024; Accepted 21-Mar, 2024; to appear in ApJ)
\end{center}

\begin{abstract}
The {\em bulk flow} in the Local Universe 
is a collective phenomenon due to the peculiar motions of matter structures, which, instead of moving in random directions, appears to follow an approximate dipole velocity flow. 
We apply a directional analysis to investigate, through the Hubble-Lema\^{\i}tre diagram, the angular dependence of the Hubble constant $H_0$ of a sample of Type Ia Supernovae from the Pantheon+ catalog in the Local Universe ($0.015 \le z \le 0.06$). 
We perform a directional analysis that reveals a statistically significant dipole variation of $H_0$, at more than $99.9\%$ confidence level, showing that matter structures follow a dipole bulk flow motion towards $(l,b) = (326.^\circ1 \pm 11.^\circ2,27.^\circ8 \pm 11.^\circ2)$, close to the Shapley supercluster 
$(l_{\scalebox{0.6}{Shapley}},b_{\scalebox{0.6}{Shapley}}) = (311.^\circ5, 32.^\circ3)$, with velocity $132.14 \pm 109.3$ km s$^{-1}$ at the effective distance $102.83 \pm 10.2$~Mpc. 
Interestingly, the antipodal direction of this dipole points close to the Dipole Repeller structure.  Our analyses confirm that the gravitational dipole system Shapley-Dipole Repeller explains well the observed bulk flow velocity field in the Local Universe. Furthermore, we performed robustness tests that support our results. Additionally, our approach provides a measurement of 
the Hubble constant $H_0 = 70.39 \pm 1.4$~\text{km s$^{-1}$ Mpc$^{-1}$}, 
at the effective distance $102.8$~Mpc, $z \simeq 0.025$. Note that this value was obtained using the first order approximation of the Hubble law because our methodology is model-independent. If one assumes, for instance, cosmography at second order with the $\Lambda$CDM value $q_0 = -0.55$, which is a model-dependent hypothesis, then $H_0 = 72.6 \pm 1.5$ km s$^{-1}$ Mpc$^{-1}$, but our results: bulk flow velocity, dipole direction and its statistical significance remain the same.

\end{abstract}



\begin{multicols}{2}

\section{Introduction} \label{sec:intro}

Our Local Universe is manifestly inhomogeneous, plenty of matter structures, like galaxies assembled in small groups or in galaxy clusters, and cosmic voids, with various underdense matter 
contents and sizes, around us \citep{rubin1951, gerard1953, gregory1978, lapparent1986, Tully87, Petercoles1996, Courtois13, Tully19, Avila19, Franco24}. 
Peculiar velocities of cosmic objects are the result of gravitational fields, which, in turn, are originated by the surrounding matter distribution \citep{Peebles1980, Kaiser87}. 
Therefore, peculiar motions are one of the best tracers of matter density fluctuations in the universe. 

Near the Local Group of galaxies, to which the Milky Way belongs, a large underdensity termed the {\em Dipole Repeller} (DR) was recently discovered \citep{Hoffman2017}. 
Another prominent feature in our cosmic neighborhood is the {\em Shapley} supercluster \citep{Raychaudhury1989,Scaramella1989}, the largest cluster of galaxies of the Local Universe. 

The system Shapley-DR acts, approximately, as a {\em gravitational dipole system} producing a bulk flow motion of matter in the Local Universe \citep{Hoffman2017}. 
In this scenario the DR acts as it were a repulsive body causing the evacuation of matter in its surroundings; on the other side, the Shapley supercluster acts as the dominant matter attractor, pulling matter structures from every side. 
One important motivation for studying this gravitational dipole system, and the features of the induced bulk flow, is due to the 
dynamical effects produced as large peculiar velocities in our nearby galaxies \citep{Tully19,Peterson22}, 
information needed to a better calibration of the standard candles for precise measurements of $H_0$ at low redshifts \citep{Scolnic_2018}.

Measurements resulting from some earlier studies already suggested a large bulk flow \citep{rubin1976, lynden1988}. 
More recently, reports in the literature (see, e.g.,~\cite{Hong14,Marinoni2023, McConville2023, Perivolaropoulos23}) have studied the magnitude and direction of this bulk flow velocity with diverse approaches and investigating various cosmic tracers, where these studies confirm the dipole nature of the bulk flow, but point out some differences in the velocity magnitude and/or direction \citep{TURNBULL2012, Scrimgeour2016, qin18, Avila23, Watkins23, Whitford23}. 

In this work, we perform an accurate directional analysis of the Type Ia Supernovae (SNe Ia) data from Pantheon$+$ \citep{pantheonshoes} to measure the bulk flow velocity, both in intensity and in direction. 
Our analyses differ from other studies in various aspects. 
First, our methodology follows previous studies looking for preferred directions on the sky using several cosmic tracers \citep{Bernui08,Marques18,Kester23}; 
second, to avoid systematics from different observables, we use only SNe Ia data from Pantheon+; this choice is a challenge due to the possibility 
of having few SNe in some directions of the sky; 
third, we measure $H_0$ and its uncertainty in a set of $N$ directions, which covers the celestial sphere, performing a best fit procedure using the Pantheon+ catalog and its covariance matrix, without cosmological model assumptions; 
fourth, we perform consistency and robustness tests to support our analyses and results. 

Our directional analysis confirms an approximate dipole behaviour of the Hubble constant, 
with a dipole direction close to the Shapley-DR direction. 
The statistical significance is obtained by comparison with a set of simulated maps, 
produced by randomizing the angular positions of the SNe sample in study, and repeating our directional analysis procedure.  For consistency, our analyses scan the celestial sphere considering two angular 
resolutions, but our final results and conclusions are drawn from the best angular resolution, i.e., 
scanning the sky with $N=192$ ($N_{\text{side}} = 4$) spherical caps. 

The outline of this work is the following. 
In Section~\ref{sec:data} we select the SNe data, subsample of the Pantheon+ SNe Ia catalog, for our directional analysis in the Local Universe. 
In Section~\ref{sec:methodology} we describe in detail the directional analysis approach, including the measurements and uncertainties involved. 
In Section~\ref{sec:results}, we present the $H_0$-maps, calculate their dipole component, and determine its statistical significance. 
In Section~\ref{sec:final} we present our conclusions and final remarks. 
We leave for the Appendix our consistency and robustness tests that support our analyses.

\vspace{0.0cm}


\section{Data: The Pantheon+ SN\MakeLowercase{e} I\MakeLowercase{a} Catalog}
\label{sec:data}

\begin{figure*}[!ht]
\centering
\includegraphics[width=0.46\textwidth]{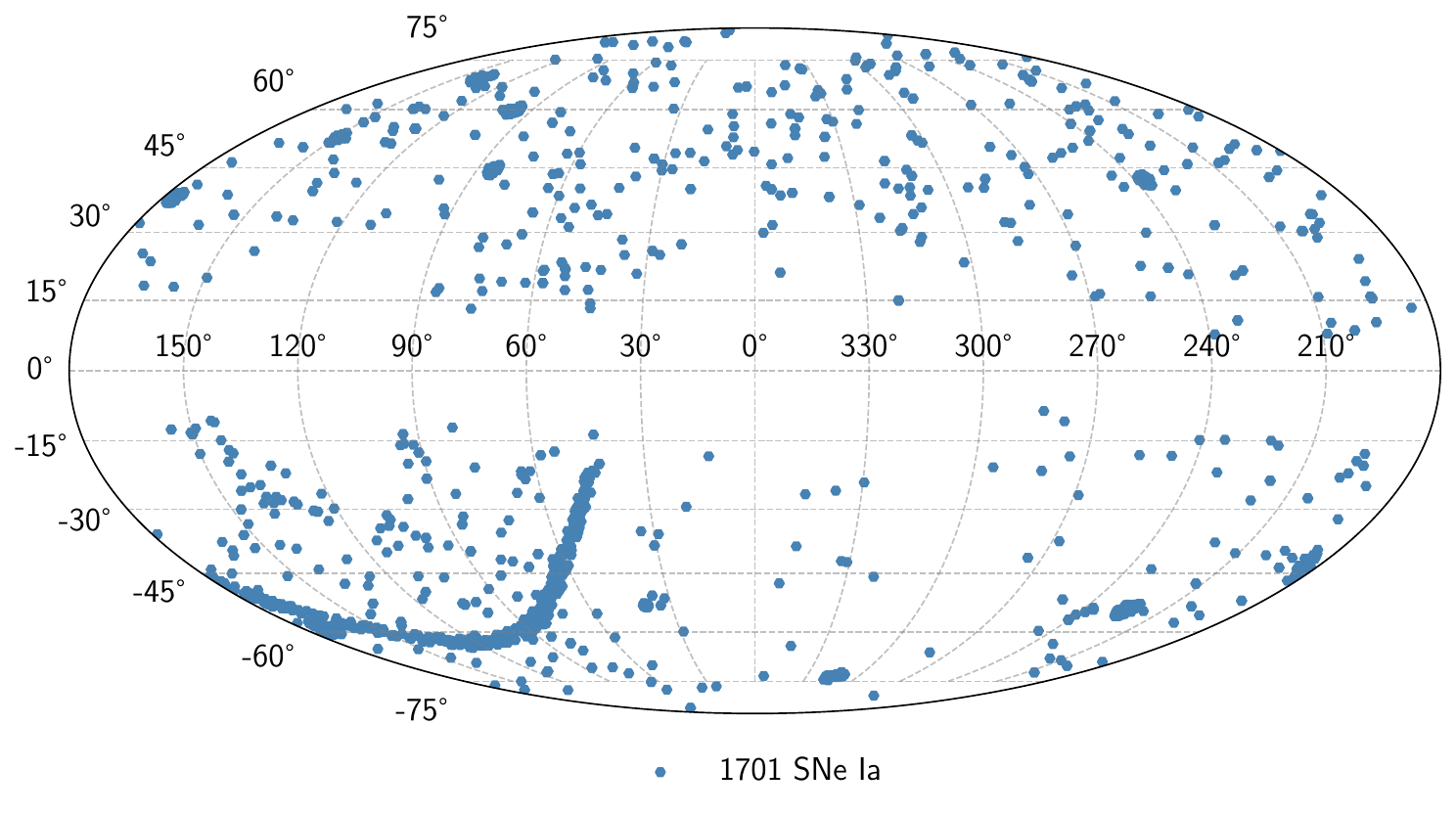}  \quad
\includegraphics[width=0.46\textwidth]{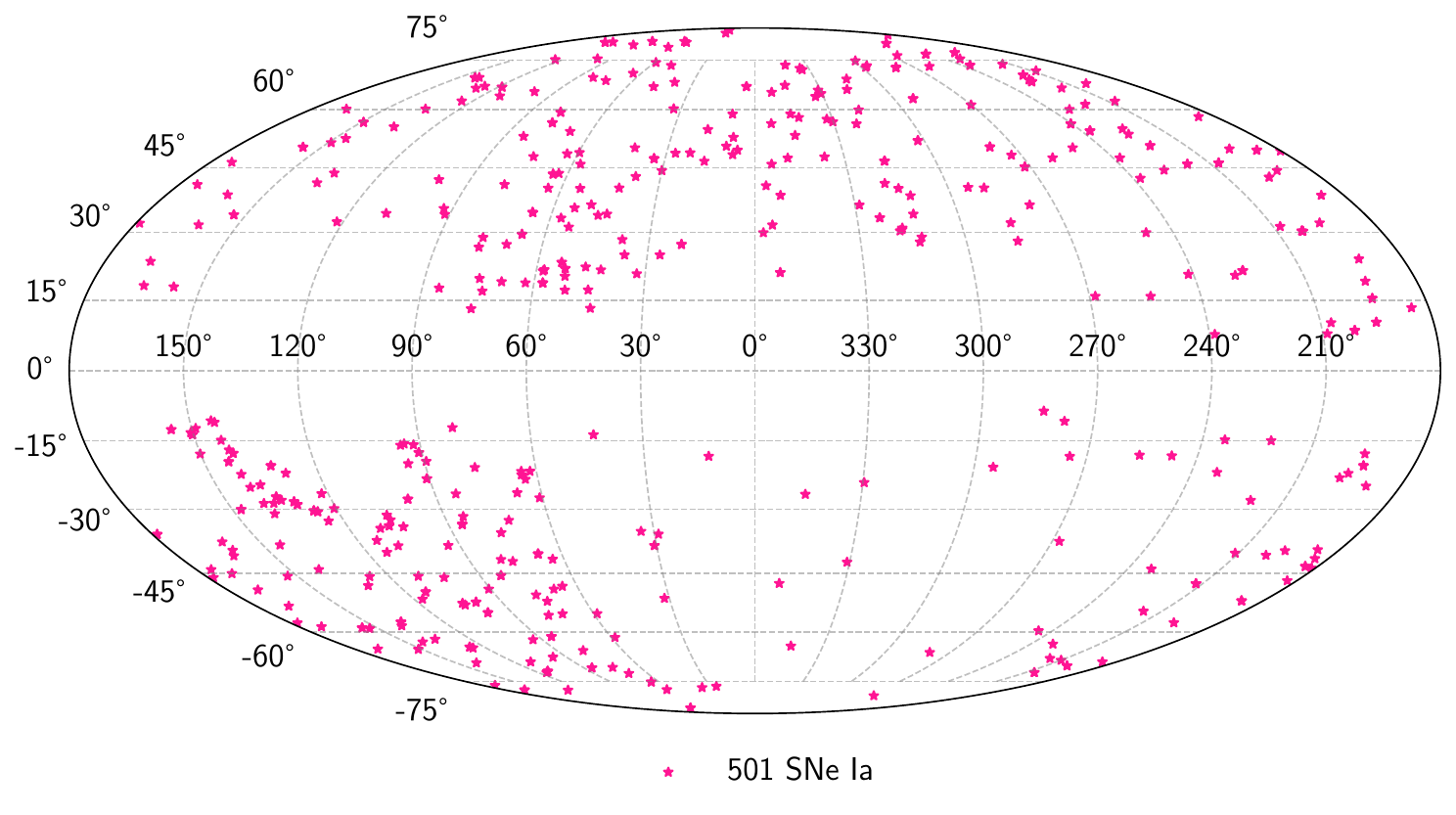}
\caption{Distribution of SNe Ia on the celestial sphere with a Mollweide projection in galactic coordinates. 
{\bf Left panel:} the Pantheon$+$ sample, showing $1701$ SNe Ia. 
{\bf Right panel:} the sample of 501 SNe Ia selected for our directional analysis, within the redshift interval $0.015 \leq z_{\scalebox{0.6}{CMB}} \leq 0.06$.}
\label{figure_pantheon+}
\end{figure*}

The recently released catalog of SNe Ia {\em Pantheon+} \citep{pantheonshoes}, successor of the original {\em Pantheon} sample \citep{Scolnic_2018}, compiles a larger number of SNe Ia events, with more precise data, and it includes events located in host galaxies whose distances were obtained using {\em Cepheids}. The Pantheon+ catalog contains $1701$ type Ia SNe, with redshifts $0.001 \leq z \leq 2.26$, and their angular distribution is shown in the top panel of Figure~\ref{figure_pantheon+} (in galactic coordinates, where the Milky Way is in the center of the figure. 

In addition to encompassing several redshift reference frames, such as the cosmic microwave background (CMB), and Heliocentric, the Pantheon+ catalog offers a comprehensive collection of precise data. 
This includes the light curves, covariance matrix, distance modulus, bolometric magnitude, and their associated uncertainties \citep{Scolnic2022ApJ...938..113S}. 
In our analyses, we shall use the following information from SN Ia data: 
the sky angular position, the redshift, the distance modulus, the peculiar velocities, all with their respective uncertainties except the last one, and the covariance matrix. 
Regarding the peculiar velocities' uncertainties, it is worth mentioning that they are undetermined (in the catalog appears the value $250$ ~km~s$^{-1}$ for all the SNe listed there). 

Our objective in this work is to detect the bulk flow motion in the Local Universe investigating the latest SNe Ia data from Pantheon+. 
The choice of the redshift interval for analyses was motivated by the following: 
(i)~the gravitational dipole in analysis is limited by the region spatially close to the Shapley supercluster, where $z_{\text{Shapley}} \simeq 0.05$; for this, we consider data with $z_{\scalebox{0.5}{CMB}} \le 0.06$; 
(ii)~galaxies very close to us are not expected to follow the Hubble flow, as overdense and underdense  regions in our neighborhood affect their motion, making it dominant over the Hubble expansion, 
then we adopt a conservative low-redshift limit 
$z_{\scalebox{0.5}{CMB}} \ge 0.015$~\citep{Riess2016,Peterson22}. 
In this case, our analyses are restricted to the Local Universe; therefore one can use the Hubble-Lema\^{\i}tre (HL) law to find $H_0$ \citep{Visser2004}. 
Notice that, for consistency, in Appendix~\ref{ap:redshift-intervals} we consider other four redshift intervals, close to $z_{\scalebox{0.6}{CMB}} \in [0.015, 0.06]$, where we apply our directional analysis to study the SNe samples contained in these intervals; as observed there, we basically obtain the same results. 

\begin{figure*}[!ht]
\centering
\includegraphics[width=0.5\linewidth]{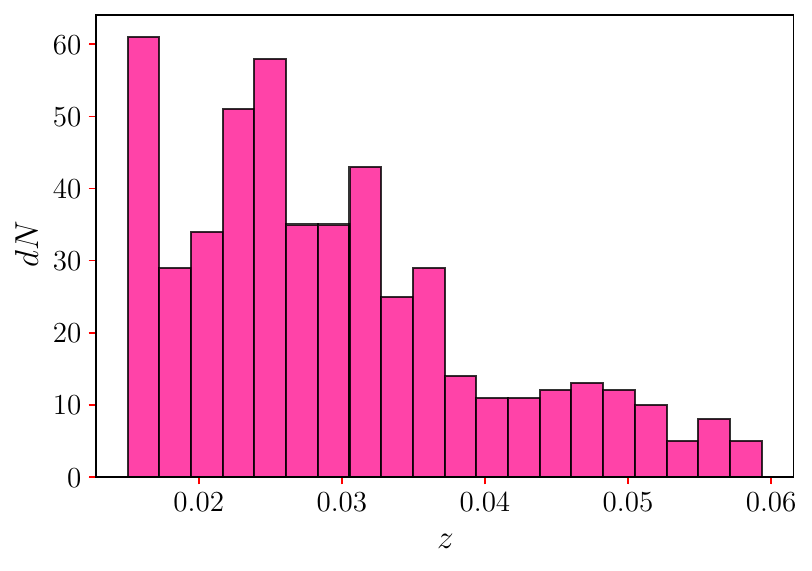}
\caption{Frequency histogram of the 501 SNe Ia sample selected for our directional analysis, in different redshift intervals within the range $0.015 \leq z_{\scalebox{0.6}{CMB}} \leq 0.06$; the bin size is $\Delta z = 0.002$.}
\label{hist_redshift_local}
\end{figure*}

According to these selection criteria, the sample chosen for analyses has $501$ SNe Ia with redshifts  $0.015~\leq~z_{\scalebox{0.6}{CMB}}~\leq~0.06$, i.e., measured in the CMB reference frame. 
The angular distribution of this sample is shown in galactic coordinates in the lower panel of Figure~\ref{figure_pantheon+}, and its redshift distribution is given in Figure~\ref{hist_redshift_local}. 
\noindent


\section{Methodology}
\label{sec:methodology}

In this section, we present our methodology to perform the directional analysis of the set of 501 SNe Ia selected in the previous section from the Pantheon+ catalog. 
Our study consists in determining the value of $H_0$ in different sky directions. 
For this we define a set of $N$ directions, chosen according to the HEALpix scheme, 
that uniformly scan the celestial sphere with spherical caps of angular radius 
$\gamma$ centered at these $N$ directions. 
Our directional analysis considers two angular resolutions to scan the celestial sphere: 
with $N=48$ ($N_{\text{side}} = 2$) and $N=192$  ($N_{\text{side}} = 4$) spherical caps. 
Then, we perform the best fit analysis in the HL diagram for those SNe Ia inside the $i$th spherical cap, centered at direction $i$, obtaining the best fit value $H_0^{b\text{-}\!f} \rightarrow H_0^{i}$ at the  $i$th direction. 
We repeat this procedure for the $N$ directions, obtaining the data set~$\{ H_0^{i} \}$, for $i=1,2,\cdots,N$. 
The details of this analysis will be discussed in Section~\ref{sec:H0maps}. 

\subsection{Finding \texorpdfstring{$H_0$}{} with the Hubble-Lemaître Law}

In the CMB reference frame, the measured velocity of any cosmic object is \citep{Courteau99,Baumann22}
\begin{equation}
\vec{\text{v}} = H_0 \,\vec{r} +  \vec{\text{v}}_{pec} \,,
\end{equation}
where peculiar motions contribute to deviate cosmic objects from the Hubble flow (i.e., the universe expansion). 
Considering the radial component of this equation, one obtains 
\begin{equation}\label{eq:Hubble_law}
c \,z_{\scalebox{0.6}{CMB}} = H_{0}\,D + \text{v}_{pec}^{r} \,,
\end{equation}
where $D$ is the radial physical distance to the cosmic object, 
and $\text{v}_{pec}^{r}$ is the radial projection of its peculiar velocity 
(to simplify the notation, from now on, we write: $\text{v}_{pec} = \text{v}_{pec}^{r}$). 
 
\begin{figure*}[!ht]
\centering
\includegraphics[width=0.49\linewidth]{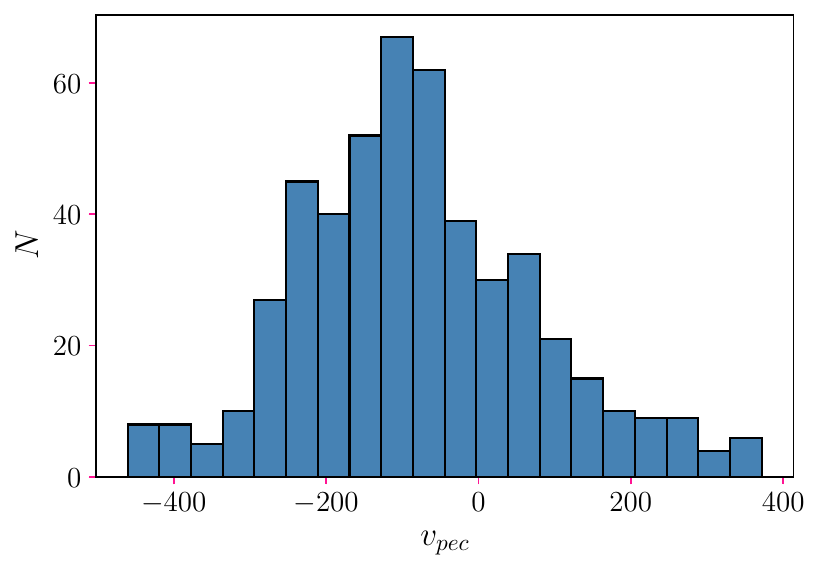}
\includegraphics[width=0.49\linewidth]{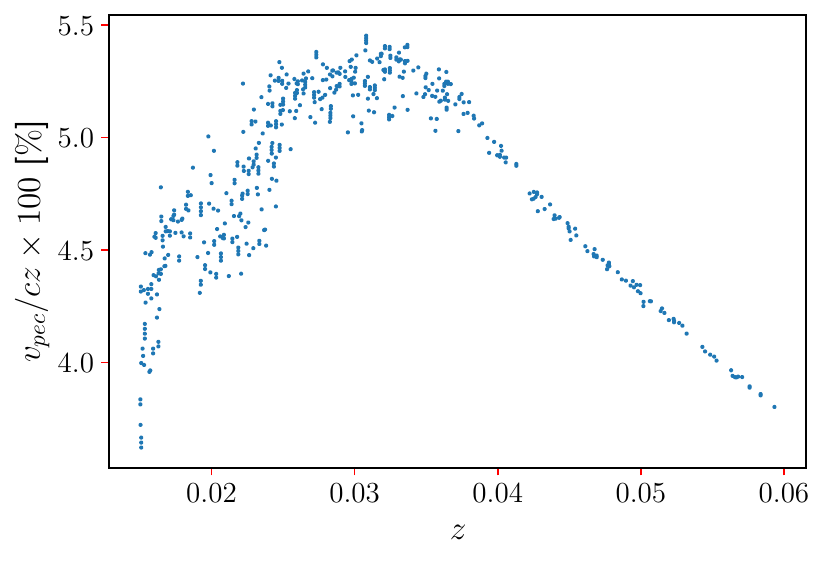}
\caption{{\bf Left panel:} peculiar velocities histogram of the 501 SNe Ia selected for our bulk flow analyses; this histogram shows that mean $-81.87$, median $-97.44$, and standard deviation $157.90$, all in units of km s$^{-1}$. 
{\bf Right panel:} fraction, in percentage, of the peculiar velocities with respect to the recession velocity for the sample in analysis.
}
\label{fig:pecvel}
\end{figure*}

The way we adopt to minimize the impact of the peculiar velocities is to choose for analysis SNe with $z_{\scalebox{0.6}{CMB}} \geq 0.015$, because in such a case, the peculiar velocities are a small fraction of the radial recession velocity. 
In fact, as observed in the plots shown in Figure~\ref{fig:pecvel} the peculiar velocities of our sample in analysis are, on average, 
 lower than 5\% of the recession velocity at redshifts 
$z_{\scalebox{0.6}{CMB}} \geq 0.015$. 
The impact of the SNe peculiar velocities will be considered later in our analyses, in Section~\ref{effdis-pecvel}. 

\subsubsection{Luminosity distance}

To estimate the value of the Hubble constant, $H_0$, through a best fit analysis, 
we use the SNe Ia luminosity distance and the redshift of each SN. 
In the Local Universe, luminosity distance is a good approximation for estimating proper distances to objects with small redshifts, $z < 0.2$~\citep{Visser2004}. 
In effect, to perform the best fit for $H_0$, we use the first-order expansion of the luminosity distance derived for the HL law, resulting from the series expansion for the physical distance $D(z) = D_L(z)$, 
\begin{equation}
D_{L}(z_{\scalebox{0.5}{CMB}}) = \frac{c \,z_{\scalebox{0.5}{CMB}}}{H_0},
\end{equation}
where $z_{\scalebox{0.5}{CMB}}$ is the SN redshift in the CMB frame. 

To calculate the SNe Ia luminosity distances, we use the distance modulus $\mu$, where the fiducial magnitude was determined from SNe host distances \citep{pantheonshoes}. 
The distance modulus is defined as the difference between the apparent magnitude $m_B$ and the absolute magnitude $M_B$ \citep{galaxy2007} 
\begin{equation}
\mu \equiv  m_B - M_B = 5\log_{10}(D_L/\text{Mpc})+25 \,,
\end{equation}
where $D_L$ is given in $\text{Mpc}$. 
Then, we calculate the luminosity distance as 
\begin{equation} 
D_L = 10^{\frac{\mu - 25}{5}} \,. 
\end{equation}


\subsubsection{Uncertainties associated to \texorpdfstring{$D_L$}{}  
and \texorpdfstring{$\delta H_{0}$}{}}\label{uncertainties}

To determine the $1 \sigma$ uncertainty of a function $X$ 
that depends on two or more variables $u, v, ...$, $X = f(u, v, ...)$, we calculate it by quadrature 
\begin{equation}
\sigma_{_X} = 
\sqrt{\left(\frac{\partial X}{\partial u}\right)^2 
(\sigma_u)^2 
+ \left(\frac{\partial X}{\partial v}\right)^2 
(\sigma_v)^2 
+ \cdots} \,\,,
\label{prop_erro}
\end{equation}
where the terms $\frac{\partial X}{\partial u}$, 
$\frac{\partial X}{\partial v}, ...$, are the partial derivatives of $X$ with respect to $u, v, ...$, and $\sigma_{u}, \sigma_{v}, ...$, are the uncertainties associated with the measured quantities  $u, v, ...$, respectively. 
According to this, the errors associated to measurements of the luminosity distance, $D_L$, are
\begin{eqnarray}
\sigma_{_{D_L}}   &=& \frac{1}{5} \ln(10) \,D_L \,\sigma_{\mu} \,, 
\label{err_distance}
\end{eqnarray}
where $\sigma_{\mu}$ is directly obtained 
from the Pantheon+ catalog.
To estimate the error of $H_{0}$ we take into account 
the uncertainties from the best fit procedure, which in turn uses 
the covariance matrix of the Pantheon+ sample, and also consider the 
statistical error due to the number of SNe Ia in each spherical cap. 
Therefore, the total error in the measurement of $H_0$ is
\begin{equation}
\sigma_{H_0} = \sqrt{ \sigma_{b\text{-}\!f}^2 + \sigma_{st}^2 + \sigma_{dp}^2} \,\,,
\end{equation}
where $\sigma_{b\text{-}\!f}$ is the uncertainty in the best fit, and $\sigma_{st}$ corresponds to the statistical error calculated as
\begin{equation}
\sigma_{st}^{i} = \frac{1}{\sqrt{n_i}} \,,
\end{equation}
where $n_i$ is the number of SNe Ia observed in the spherical cap $i$, and $\sigma_{dp}$ 
corresponds to the uncertainty due to the dipole nature of the $H_0$-map (see Section~\ref{dipoleAPS}) uncertainty to be calculated using a Monte Carlo simulation in Appendix \ref{appendixA}.

\subsection{The SNe \texorpdfstring{$H_0$}{}-maps}

To investigate in what direction our sample of 501 SNe Ia, with redshifts $0.015 \le z \le 0.06$, show a larger value of $H_0$, or otherwise does not show a preferred direction, we perform our directional analysis. 
For this, we shall scan the celestial sphere with $N$ spherical caps, of radius $\gamma$, and consider two cases: $N = 48$ and $N = 192$ caps. 

Let $\Omega_{i}^{\gamma} \equiv \Omega(\theta_{i}, \phi_{i}, \gamma) \in \mathcal{S}^{2}$ be a spherical cap on the celestial sphere, 
of $\gamma$ degrees of aperture, with vertex at the $i$th pixel with coordinates $(\theta_i, \phi_i)$, 
that is, 
$(\theta^i_{\text{cap}}, \phi^i_{\text{cap}}) = (\theta_i, \phi_i)$, for 
$i = 1,2,\cdots, N_{\text{pix}}$ (for details of our directional analysis see~\cite{Bernui08}). 
Both, the number of spherical caps $N$ and the angular coordinates of their centers $\{ (\theta^i_{\text{cap}}, \phi^i_{\text{cap}}) \},\, i = 1,2,\cdots, N$, are defined using the 
Healpy~HEALPix\footnote{\url{https://healpy.readthedocs.io/en/latest/}} pixelization algorithm \citep{healpix2005ApJ...622..759G}.

The pixelization parameter $N_{\text{side}}$ provides the grid resolution that defines the quantity of pixels, $N_{\text{pix}}$, in which the sphere is pixelized
\begin{equation}
N = N_{\text{pix}} = 12 \, N_{\text{side}}^{2} \,, 
\end{equation} 
where $N_{\text{side}} = 2^{k}$, and $k$ is an integer number sometimes called 
{\it order}. 
This means that $N_{\text{side}} = 2$ produces $N = 48$ cap centres, and  $N_{\text{side}} = 4$ produces $N = 192$ cap centers, which correspond to the angular resolutions we will use in our directional analysis. 

As observed in the lower plot of Figure~\ref{figure_pantheon+}, the SNe are not uniformly distributed on the sky, 
this is because the Pantheon+ catalog is a compilation of SNe that are short-lived random events in the sky, not long-lived cosmic tracers that can be observed at any time with an astronomical survey. 
This implies that the number of SNe is not equal in all caps, a fact that we take into account in our error analysis, and in the choice of $\gamma$, the radius of the spherical caps.

\subsubsection{From the \texorpdfstring{$H_0$}{}-caps to the \texorpdfstring{$H_0$}{}-map}\label{sec:H0maps}

Our directional analysis consists in calculating $H_0$ in each one of the $N$ caps that scan the celestial sphere, computation that is done performing the best fit of the HL diagram for the SNe observed in each cap, and assigning the $H_0$ value found to the centre of the cap. 
For example, suppose that in the $i$-th spherical cap, of radius $\gamma$, we observe $n_i$ SNe, then we perform the best fit of the HL diagram with these $n_i$ events finding the value $H_0^i$. 
In Appendix \ref{pearson}, we investigate the correlation between the number of SNe and the best fit $H_0$, where we conclude that the number of SNe observed in each of the spherical caps has a  negligible impact in the dipolar behaviour of $H_0$. 
The set of $N$ values: $\{ H_0^i \}$, for $i = 1, \cdots, N$, are then 
assembled together into a full-sky map, hereafter the $H_0$-map; 
attributing a color scale to these set of real numbers $\{ H_0^i \}$ 
we obtain a colored $H_0$-map, as seen in the left panels of  Figure~\ref{fig:H0maps}. 
The next step is to calculate the dipole component of the $H_0$-map, and establish its 
statistical significance with the help of simulated isotropic maps (see 
Section~\ref{isotropic-maps}). 

The radius of the spherical cap, $\gamma$, is a challenge: 
it cannot be small because the number of SNe observed in that caps would be so small that the best fit is dominated by statistical noise, 
and cannot be large that one loses precision\footnote{Because including SNe that do not contribute to the direction of interest causes the decrease of  the effect we want to measure.} in determining the directions where $H_0$ is maximum or minimum. 
As a lower limit for the number $n_i$ we consider the number of the 
SNe analyzed in 1998 \citep{Riess1998,Perlmutter1998}, that is $n_i \sim 40$ SNe. 
After various tests, we adopt $\gamma = 60^\circ$. 
However, we also performed robustness tests with other cap radius to verify that the value $\gamma = 60^\circ$ is not biasing our results. 
As shown in Appendix~\ref{othercaps}, we redo our directional analysis considering 
caps with $\gamma = 65^\circ$ and $\gamma = 70^\circ$ obtaining completely similar results.


\section{Analyses and Results}
\label{sec:results}

In this section, we shall present our analyses and corresponding results. 
We start calculating the $H_0$-maps, for the two angular resolutions, with 
$N = 48$ and $N = 192$ spherical caps. 
Then, we calculate the dipole components of these maps and determine their statistical significance. 


\subsection{Dipole structure and the angular power spectrum}
\label{dipoleAPS}

\begin{figure*}[t]
\begin{center}
    \includegraphics[width=0.45\linewidth]{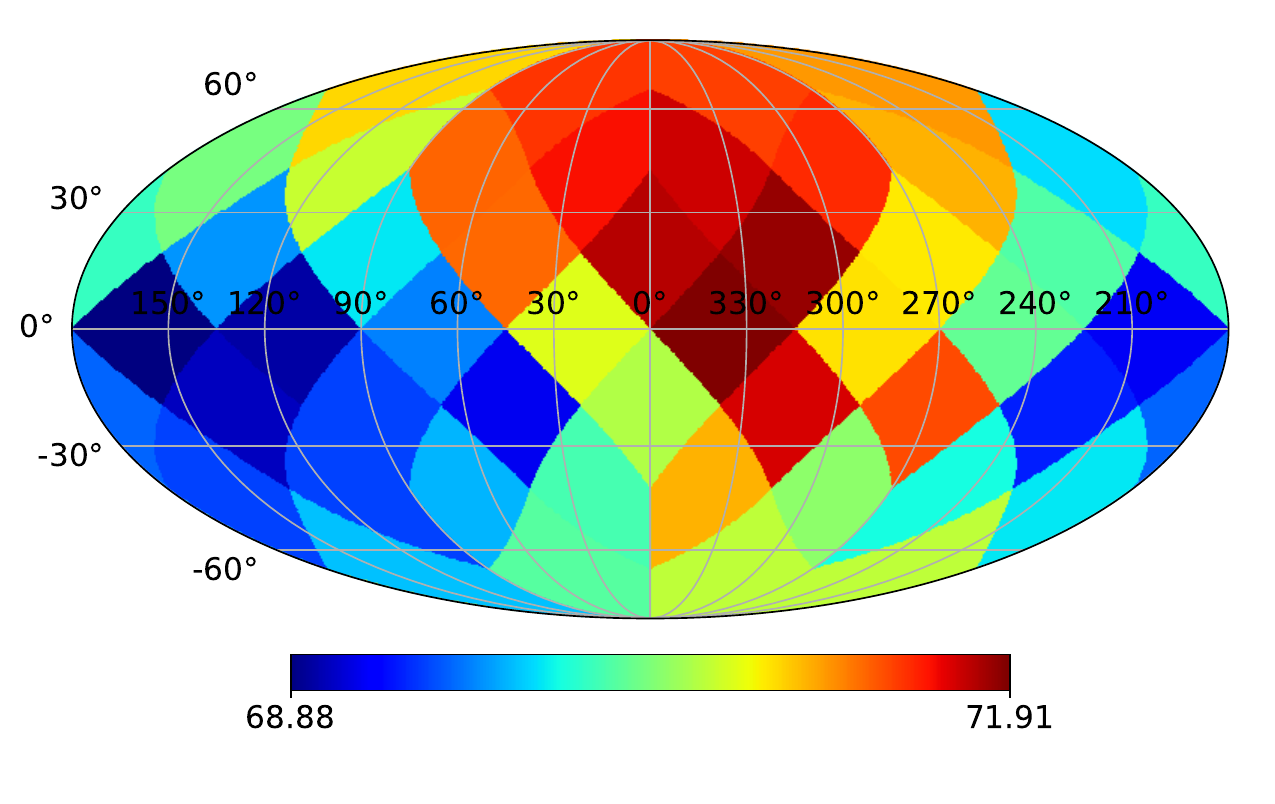}\includegraphics[width=0.45\linewidth]{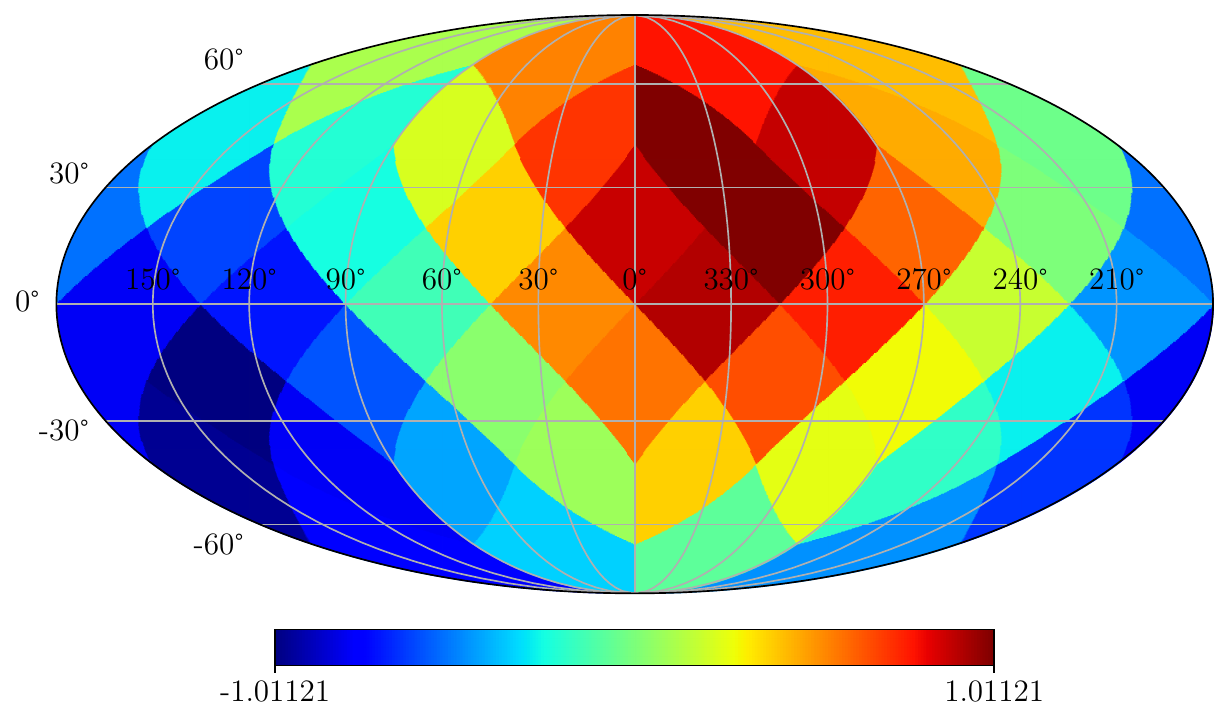} \\
    \includegraphics[width=0.45\linewidth]{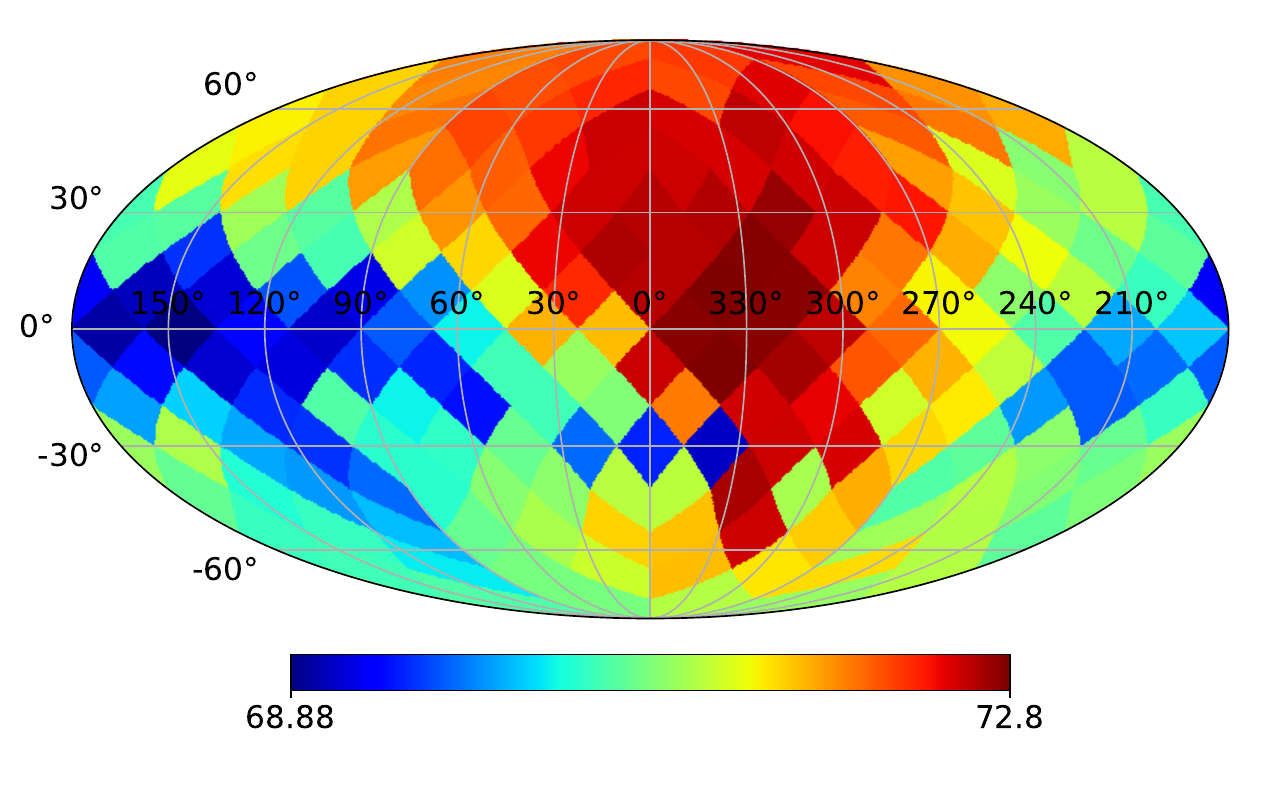}\includegraphics[width=0.45\linewidth]{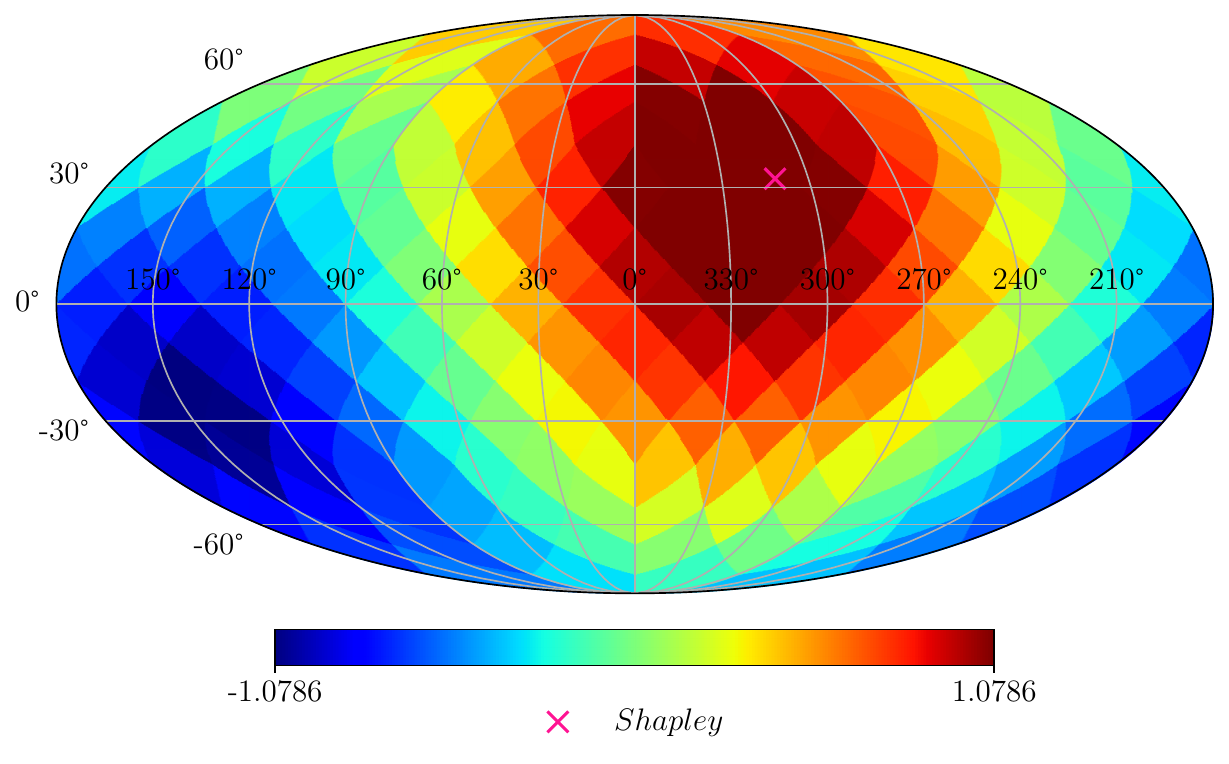}
\caption{
{\bf First row:} 
the $H_0$-map$^{48}$ with resolution $N_{\text{side}}=2$ (48 spherical caps, left panel), and its dipole component map (right panel).
{\bf Second row:} 
the $H_0$-map$^{192}$ with resolution $N_{\text{side}}=4$ (192 spherical caps, left panel), and its dipole component map 
(right panel) showing the position of the Shapley supercluster as a multiplication sign \textsf{X} (in fuchsia color) to evidence that the dipole direction obtained in our analyses is very close to this matter structure. 
Note that all these maps are in units km~s$^{-1}$~Mpc$^{-1}$.}
\label{fig:H0maps}
\end{center}
\end{figure*}

With all the data in hand, we then calculate the $H_0$-maps, scanning the celestial sphere with spherical caps of radius 
$\gamma = 60^\circ$. 
Our results are displayed in Figure~\ref{fig:H0maps}. 
The next step is to calculate the dipole component $C_1$ of the corresponding angular power spectra. 
Clearly, the dipole term $C_{1}$ is the larger one, $C_1^{48}=1.52$ for the $H_0$-map$^{48}$ and {$C_1^{192}=1.66$} for the $H_0$-map$^{192}$, indicating that the $H_0$-maps show a net dipolar behavior. 
However, one needs to evaluate the statistical significance of this result, and this is done by comparison with $H_0$-maps obtained by redoing our directional analysis procedure but randomizing the choice of the SNe. 
The production of these sets of isotropic $H_0^{\text{Iso}}$-maps 
considering $48$ and $192$ caps, and the statistical analyses to determine the statistical significance of the dipole components of the $H_0$-map$^{48}$ and $H_0$-map$^{192}$ are described in Section~\ref{isotropic-maps}. 

We also calculate the direction of the dipole term finding, in galactic coordinates, 
$(l,b)=(327.^\circ08 \pm 22.^\circ5, 28.^\circ23 \pm 22.^\circ5)$, 
for the $H_0$-map with $N=48$ caps, and 
$(l,b)=(326.^\circ06 \pm 11.^\circ2, 27.^\circ79 \pm 11.^\circ2)$, 
for the $H_0$-map with $N=192$ caps. 
The direction and bulk flow velocity of the dipoles, in the directional analysis with 48 and 192 spherical caps, are summarized in Table~\ref{results_table} (see also Figures~\ref{fig:H0maps},~\ref{bulkflow}, and~\ref{resultsky}).

Regarding the dipole interpretation of our $H_0$-maps, we observe 
in Figure~\ref{fig:H0maps} that this is a good approximation, but 
not an exact result. 
In fact, the color of each pixel in our $H_0$-maps represents the dominance of inflows or outflows, phenomena that increases or decreases the recession speed of matter structures, respectively, relative to the Hubble flow. 
That is, the diversity of colors in the pixels of a $H_0$-map reflects the dominance of one phenomenon or the other, or even the balance between both, effects that are quantified by the best fit $H_0^i$ from the HL diagram of those SNe in the $i$th cap: 
reddish (bluish) pixels represent $H_0^i$ values above (below) the mean, and greenish for values close to the mean.

Remarkably, the prevalence of each of these phenomena is determined by the large-scale distribution of clustered matter and large voids, and the Local Universe is plenty of 
them~\citep{Courtois13,Hoffman2017,Tully23,Dias23}. 
Large matter structures besides the Shapley supercluster, like the South Pole Wall~\cite[SPW,][]{Pomarede20}, a huge structure at $z \simeq 0.04$, produces an inflow that contributes to distort a pure dipolar pattern.
On the other hand, the presence of large underdense regions besides the 
DR \citep{Franco24}, like the Perseus-Pisces void ($z \simeq 0.027$), 
is producing outflows that also leave their peculiar signature in our 
$H_0$-maps. 
All these complex gravitational interactions produce inflows and outflows that affect the dynamics of the SN hosts, phenomena that are captured by our directional analysis and revealed in our $H_0$-maps. 
The observed dipole structure of the $H_0$-maps is just the consequence of a dominant gravitational dipole system, i.e., Shapley-DR~\citep{Hoffman2017}.

\begin{table*}[t]
\centering
\scalebox{1}{
\begin{tabular}{lcccc}
\hline
 & $l (^\circ )$ & $b (^\circ )$ & $V_{\scalebox{0.7}{BF}}$ (km s$^{-1}$) & $\delta H_{0}$ (km s$^{-1}$Mpc$^{-1}$) \\
\hline
$H_0$-map$^{48}$  & $327.08 \pm 22.5$  & $28.23 \pm 22.5$ & $127.69 \pm 110.97$  & 
$2.44 \pm 1.88$ \\

$H_0$-map$^{192}$ & $326.06 \pm 11.2$ & $27.79 \pm 11.2$ & $132.14 \pm 109.30$ & 
$2.57 \pm 1.87$ \\
\hline
\end{tabular}
}
\caption{Bulk flow direction and velocity, for directional analysis
with 48 and 192 spherical caps of radius $\gamma = 60^\circ$.}
\label{results_table}
\end{table*} 

\subsection{Bulk flow velocity}

Given a data set of low-redshift SNe~Ia, 
$z \ll 1$, one can perform a best fit analysis of the linear HL law and write  
\begin{equation}
\langle 
\frac{c\,z_{\scalebox{0.6}{CMB}}}{r} 
\rangle = H_0^{b\text{-}\!f} \,,
\end{equation}
where the angular parentheses mean the procedure done to obtain the best fit value $H_0^{b\text{-}\!f}$. 
Clearly, according to the left panels of Figure~\ref{fig:H0maps}, 
the $H_0$-maps, the value one obtains in the best fit analysis, $H_0^{b\text{-}\!f}$, depends on the direction in study. 
Analyzing these maps one determines the positive ($+$) and negative ($-$) 
dipole directions of the $H_0$-maps (right panels of Figure~\ref{fig:H0maps}). 
Consider the best fit value of the $H_0$-map obtained in the positive (negative) dipole direction, that is, $H_0^{+}$ ($H_0^{-}$). 
Then, the directional analysis of the SNe data in these directions, 
with effective distance $R$, states 
\begin{equation}
\langle 
\frac{c\,z_{\scalebox{0.6}{CMB}}}{r} \rangle_{\pm} 
= H_0^{\pm} \simeq H_0 \pm 
\frac{\text{V}_{\scalebox{0.6}{BF}}(R)}{R} \,,
\end{equation}
where we interpret the excess (defect) in $H_0$ as being the effect of the bulk flow motion, with velocity $\text{V}_{\scalebox{0.6}{BF}}(R)$ at the effective distance $R$, on the universe expansion rate measured by the SNe. 
As a consequence of this, one has
\begin{equation}\label{defdeltaH0}
H_0^{+} - H_0^{-} \equiv \delta H_{0}(R) \simeq \frac{2\,\text{V}_{\scalebox{0.6}{BF}}(R)}{R} \,.
\end{equation}
Therefore, 
\begin{equation}
\text{V}_{\scalebox{0.6}{BF}}(R) \simeq \frac{1}{2} R\,\,\delta H_{0}(R) \,.
\label{EQbulkflow}
\end{equation}
In what follows we will show how to obtain the effective distance $R$. 


\subsubsection{Effective distance \texorpdfstring{$R$}{} and peculiar velocities}
\label{effdis-pecvel}

To calculate the effective distance $R$ of a data sample, 
necessary to compute the bulk flow velocity, 
we use the thermal noise model 
of \,\cite{TURNBULL2012},
\begin{equation}\label{Effectivedistance}
R = \frac{\sum_{i=1}^{N_{\scalebox{0.5}{SNe}}} {D_{L}}_{i}
/\sigma_{i}^{2}}{\sum_{i=1}^{N_{\scalebox{0.5}{SNe}}} 1/\sigma_{i}^{2}} \,,
\end{equation}
where ${D_{L}}_i$ is the luminosity distance of the $i$-th SN, 
$\sigma_{i}$ is the peculiar velocity uncertainty of the $i$-th SN, and 
$N_{\scalebox{0.7}{SNe}}$ is the number of SNe in our sample. 
Unfortunately, the Pantheon$+$ catalog does not provide uncertainties of the SNe peculiar velocities 
measurements\footnote{The column of this quantity exists, but the information displayed there is $250$ ~km~s$^{-1}$ for all the SNe in the Pantheon+ catalog.}. 
Therefore, for the calculation of the effective distance, $R$, we need to estimate the set 
$\{ \sigma_{i} \} = \{ {\sigma_{\text{v}_{pec}}}_i \}$.

From equation (\ref{eq:Hubble_law}), the peculiar velocity in the radial direction is given by 
\begin{equation}
\text{v}_{pec} = c\,z_{\scalebox{0.6}{CMB}} 
- H_0\,D_L \,.
\end{equation}
To calculate the uncertainty of $\text{v}_{pec}$, $\sigma_{\text{v}_{pec}}$, 
we use equation~(\ref{prop_erro}) multiplied by a 
factor\footnote{The factor $1/\sqrt{2}$ is an average of all the possible directions of the peculiar velocity, relative to radial direction; see, e.g.,~\cite{pantheonshoes}.}
$1/\sqrt{2}$, obtaining\footnote{The uncertainty in the redshift measurements, $\sigma_{z_{\scalebox{0.5}{CMB}}}$, is negligible.} 
\begin{equation}\label{eqsigma-v-pec}
\sigma_{\text{v}_{\text{pec}}} = \frac{1}{\sqrt{2}}\sqrt{ D_{L}^{2} \sigma_{H_0}^{2} + H_{0}^{2}\sigma_{D_{L}}^{2}} \,. 
\end{equation}
We then compute $\sigma_{\text{v}_{\text{pec}}}$ using the best fit values of $H_0$ and $\sigma_{H_0}$, together with $D_{L}$ and $\sigma_{D_{L}}$, for each SN located in the spherical caps corresponding to the positive and negative dipole direction of the $H_0$-map. 
Performing error propagation, the uncertainty of $R$, $\sigma_{R}$, is calculated from the uncertainties in the luminosity distance, ${\sigma_{D_{L}}}_{i}$, and in the peculiar velocity, $\sigma_{i}$, that is, 
\begin{equation} \label{sigR}
\sigma_{R} = \frac{\sum_{i=1}^{N_{\scalebox{0.5}{SNe}}} {\sigma_{D_{L}}}_{i} / 
\sigma_{i}^{2}}{\sum_{i=1}^{N_{\scalebox{0.5}{SNe}}} 1/\sigma_{i}^{2}} \,. 
\end{equation}
We then calculate the effective distance using equation~(\ref{Effectivedistance}), and its associated uncertainty using equation~(\ref{sigR}), obtaining 
$R^{48} = 104.67 \pm 10.31$~Mpc and $R^{192} = 102.83 \pm 10.24$~Mpc, 
for $N=48$ and $N=192$ spherical caps, respectively. 
For consistency, we have also estimated the effective distance using Monte Carlo simulations, as presented in Appendix~\ref{ap:eff_dist}, obtaining values fully consistent with these ones.

\subsubsection{Bulk flow velocity with SNe Ia}

\begin{figure*} [!ht]
\begin{center}
\includegraphics[width=0.5\linewidth]{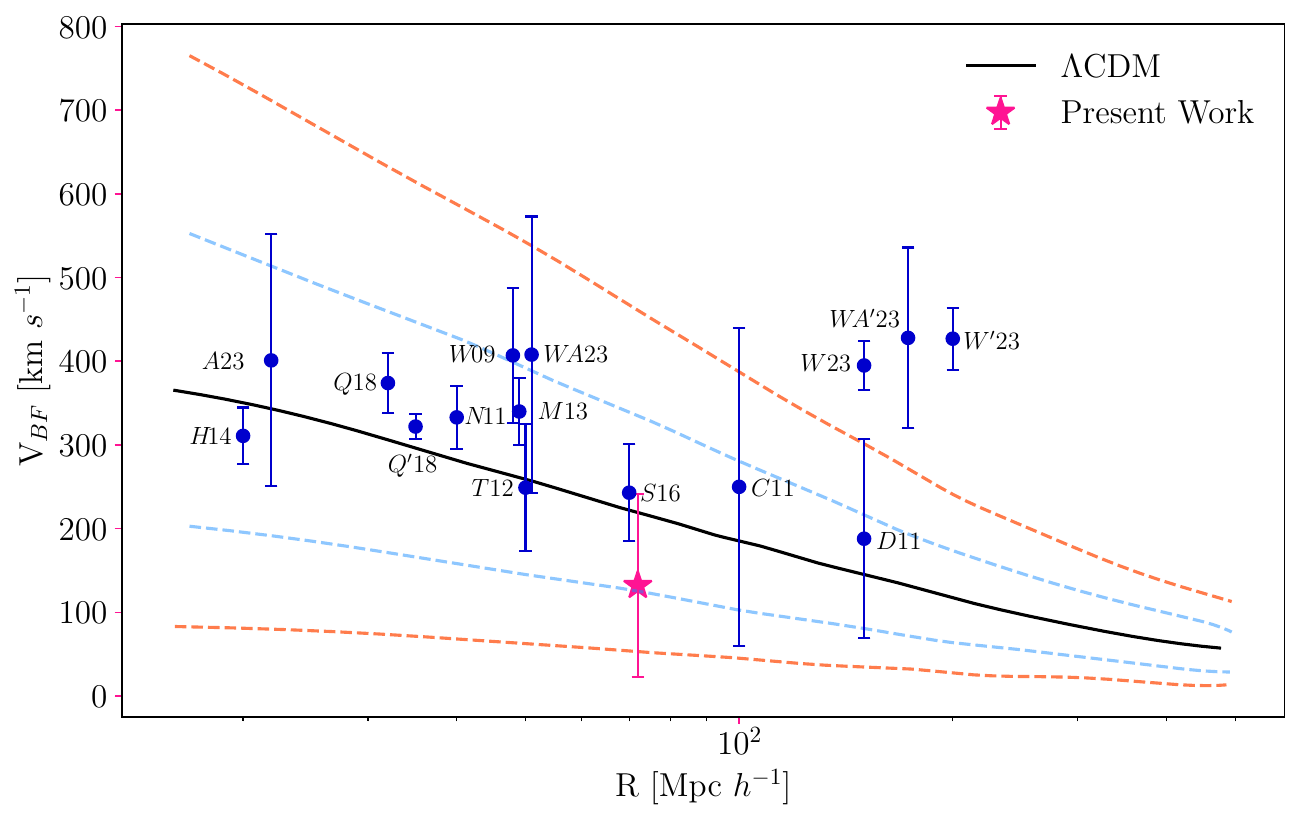} 
\caption{ This plot compares several bulk flow velocity measurements 
with the value expected in the $\Lambda$CDM model (continuous line plus $1\sigma$ and $2\sigma$ limits): A23~\citep{Avila23}; \ H14 \citep{Hong14}; \ Q18 and Q$'$18 \citep{qin18}; N11 \citep{Nusser11}; \ W09 \citep{watkins2009};\ M13 \citep{scott12}; \ T12 \citep{TURNBULL2012}; \ S16 \citep{Scrimgeour2016}; \ C11 \citep{colin11}; \ D11 \citep{Dai2011}; 
W23 and W$'$23 \citep{Watkins23}; 
WA23 and WA$'$23 \citep{Whitford23}, and the result of our analyses. All effective distances, $R$, are given in units of Mpc $h^{-1}$; in this plot, we used $h = 0.7$ for comparison purposes.}
\label{bulkflow}
\end{center}
\end{figure*}

Our measurements of the directional variation of the Hubble constant were done in Section~\ref{dipoleAPS}, where we have obtained a net dipolar structure in the $H_0$-maps, as seen in Figure~\ref{fig:H0maps}, with maximum and minimum 
values $H_{0}^{max} = 71.91 \pm 1.42$ and $H_{0}^{min} = 68.88 \pm 1.32$ 
for the $H_0^{48}$-map, and 
$H_{0}^{max} = 72.80 \pm 1.26$ and $ H_{0}^{min} = 68.88 \pm 1.17$ 
for the $H_0^{192}$-map, respectively (all $H_0$ values are in units of km~s$^{-1}$Mpc$^{-1}$). 
The differences among diverse directional measurements of $H_0$, observed in the $H_0$-maps shown in the left panels of Figure~\ref{fig:H0maps}, are interpreted as being originated in the combination of motions of the SNe host galaxies 
due to the Hubble flow (i.e., the universe expansion) and due to peculiar motions relative to the matter structures, including the bulk flow motion. 
Clearly, in a universe without peculiar motions, the $H_0$-maps should be mono-colour, indicative of the isotropy of $H_0$, that is, the same value $H_0$ in any direction (except statistical fluctuations originated in the best fit procedure of the HL diagrams). 
According to equation~(\ref{EQbulkflow}), to obtain the magnitude of the bulk flow velocity, we need $\delta H_0$. 
We calculate $\delta H_0 \equiv H_0^{+} - H_0^{-}$, 
considering the values in the $H_0$-maps (left panels of Figure~\ref{fig:H0maps}) corresponding to the $+/-$ dipole directions shown in the right panels of 
Figure~\ref{fig:H0maps}, obtaining 
$H_{0}^{+/48} = 71.78 \pm 1.34$ and 
$H_{0}^{-/48} = 69.34 \pm 1.32$, 
and 
$H_{0}^{+/192} = 72.31 \pm 1.34$ and 
$H_{0}^{-/192} = 69.74 \pm 1.30$, where all $H_0$ values are in units of km s$^{-1}$ Mpc$^{-1}$. 
The effective distances, as calculated before, are 
$R^{48} = 104.67 \pm 10.31$ Mpc and 
$R^{192} = 102.83 \pm 10.24$ Mpc, then 
\begin{equation}
V_{{\scalebox{0.6}{BF}}_{48}} 
= \frac{1}{2}(104.67 \pm 10.31) (2.44 \pm 1.88)\,,
\end{equation}
\begin{equation}
V_{{\scalebox{0.6}{BF}}_{192}} 
= \frac{1}{2}(102.83 \pm 10.24) (2.57 \pm 1.87)\,,
\end{equation}
finding $V_{{\scalebox{0.6}{BF}}_{48}} = 127.81 \pm 110.97$ km s$^{-1}$ and $V_{{\scalebox{0.6}{BF}}_{192}} = 132.14 \pm 109.30$ km s$^{-1}$, respectively.
Our results are robust, and in good agreement with the values reported in the literature (see Figure \ref{bulkflow}). 

\begin{figure*}[!ht]
\begin{center}
\includegraphics[width=0.5\linewidth]{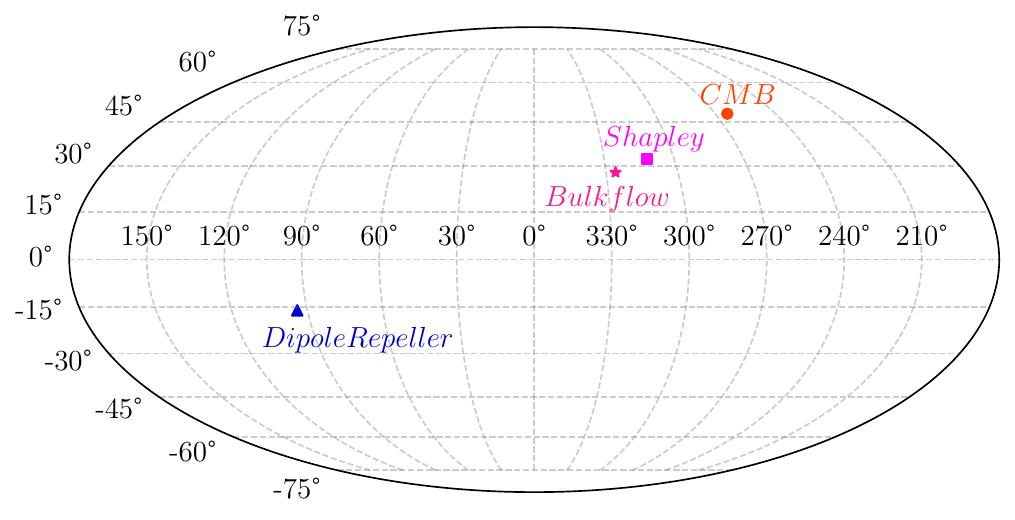}
\caption{Mollweide projection in galactic coordinates of the structures and directions characterizing flows in the Local Universe: the Dipole Repeller, Shapley supercluster, CMB dipole, and the direction of the bulk flow we found for $H_0$-map$^{192}$.}
\label{resultsky}
\end{center}
\end{figure*}

In Figure~\ref{resultsky} we can observe some interesting directions on the celestial sphere, including the dipole bulk flow motion found in our analyses.  
Our results, summarized in Table~\ref{results_table}, confirm that the gravitational system Shapley-DR explains well the dipole nature of the bulk flow motion, and that matter structures in the Local Universe 
appear to move in the direction of the Shapley supercluster with 
a velocity of $132.14 \pm 109.30$~km~s$^{-1}$.~\footnote{
The performance of our approach depends on having a large number of SNe {\it per} square degree uniformly distributed on the celestial sphere, which is not the present case; in consequence, SNe unevenly distributed on large spherical caps contribute only partially to point out the true bulk flow velocity: the potential contribution 
$\vec{\text{v}}_{pec}$ reduces to its component in the radial direction 
$\text{v}_{pec}^{r} = \vec{\text{v}}_{pec} \cdot \widehat{r}$, 
which can be a small quantity because depends on the angle between $\vec{\text{v}}_{pec}$ and the unit vector $\widehat{r}$, angle that can be large, and its {\it cosine} small, in large spherical caps. 
As a consequence, our measurement is underestimated.}

\subsection{Isotropic maps and statistical confidence analysis}\label{isotropic-maps}

We want to verify that the dipole signature found in our $H_0$-map analyses is not a biased result, perhaps due to the methodology of analysis, or some artifact in the data. 
Our statistical confidence test considers the comparison with isotropic $H_0$-maps, as follows.

\begin{figure*}[!ht]
\begin{center}
\includegraphics[width=0.4\textwidth]{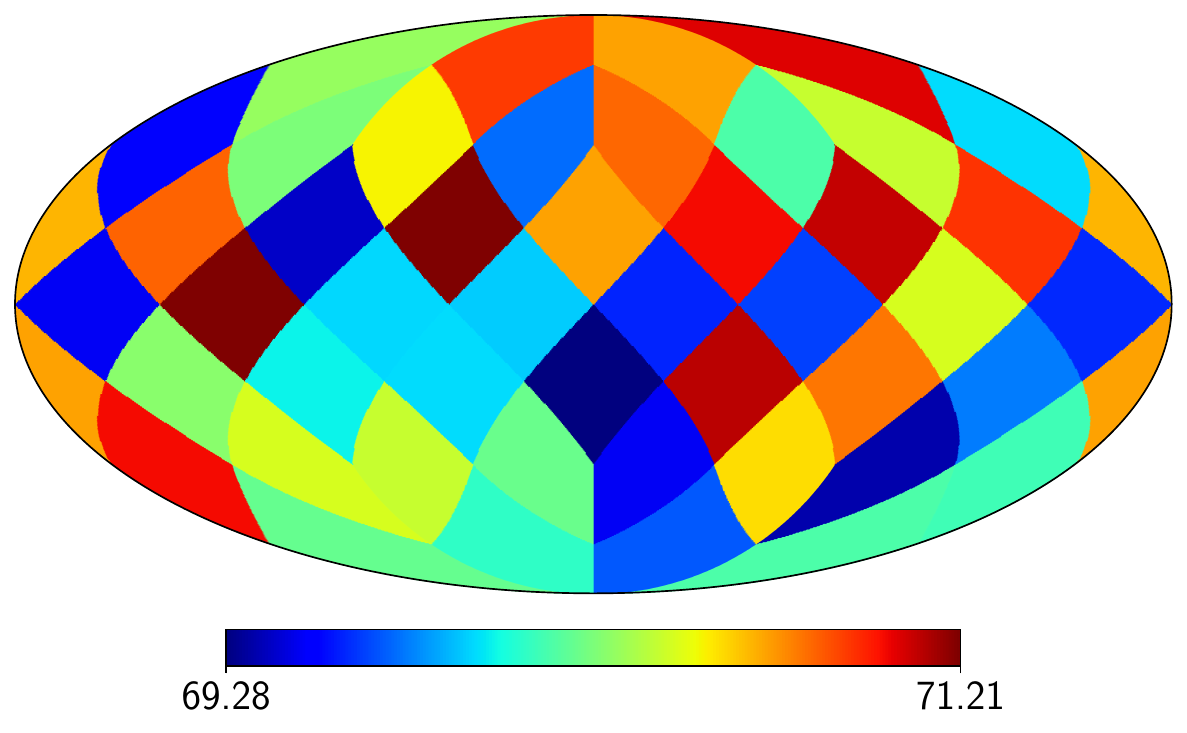}
\includegraphics[width=0.4\textwidth]{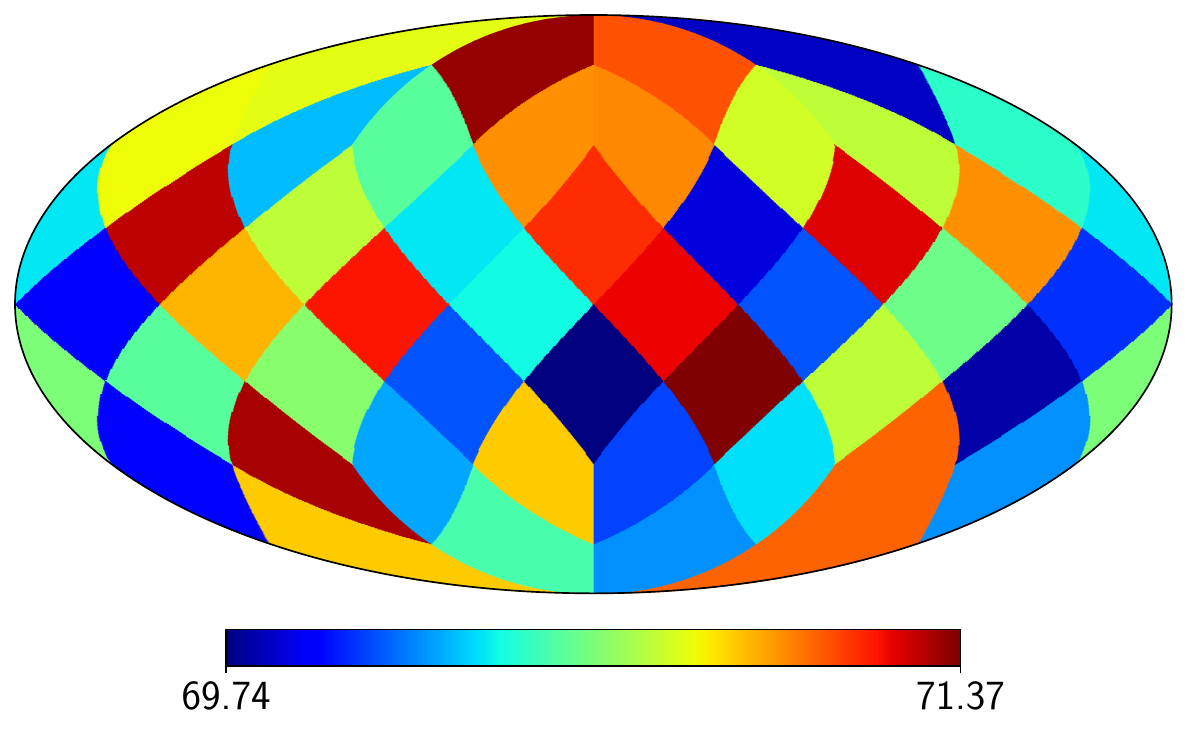} \\
\includegraphics[width=0.4\textwidth]{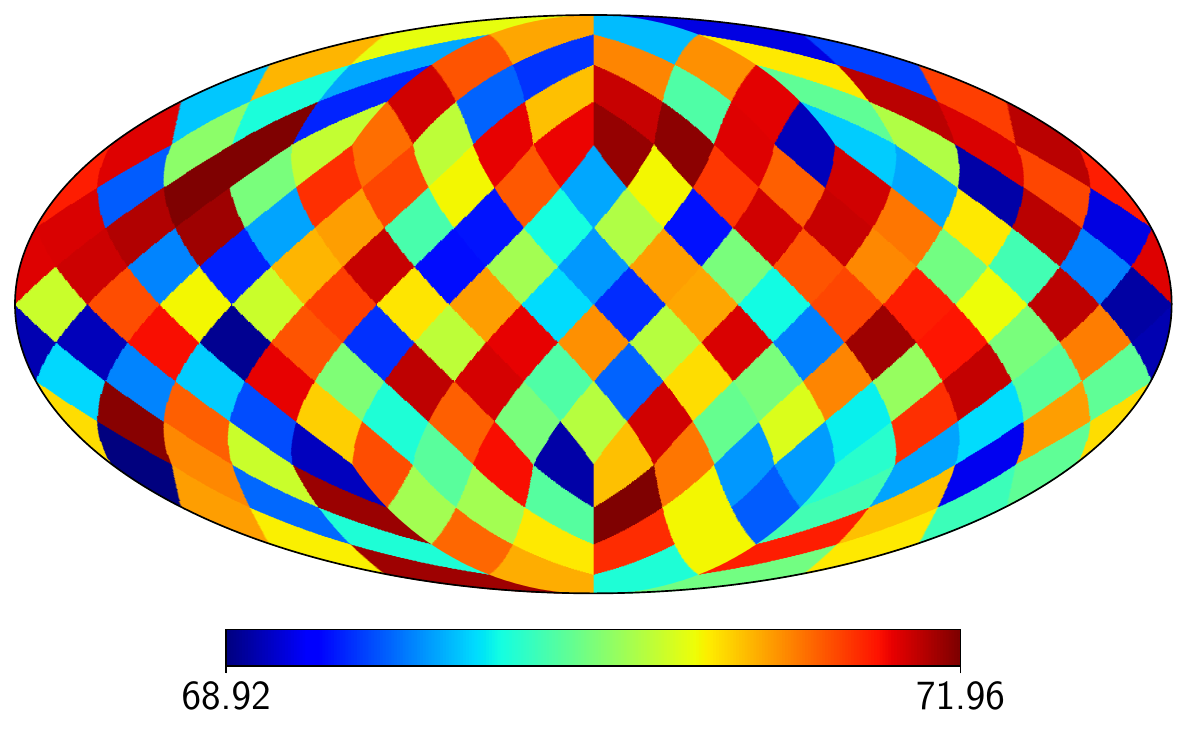}
\includegraphics[width=0.4\textwidth]{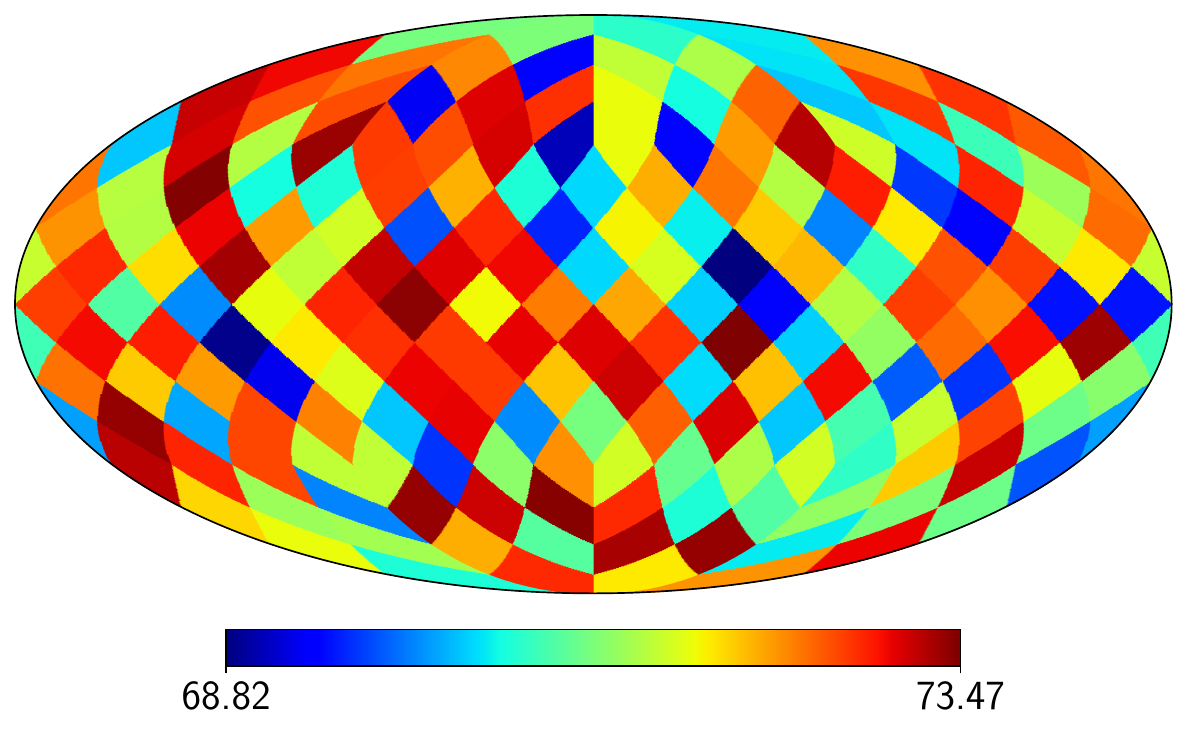}
\caption{Illustrative examples of simulated isotropic maps, $H_0^{\text{Iso-48}}$-maps, for $N=48$ caps (first row), and $H_0^{\text{Iso-192}}$-maps 
for $N=192$ caps (second row).} 
\label{maps_random}
\end{center}
\end{figure*}

\begin{figure*}[!ht]
\includegraphics[width=0.5\linewidth]{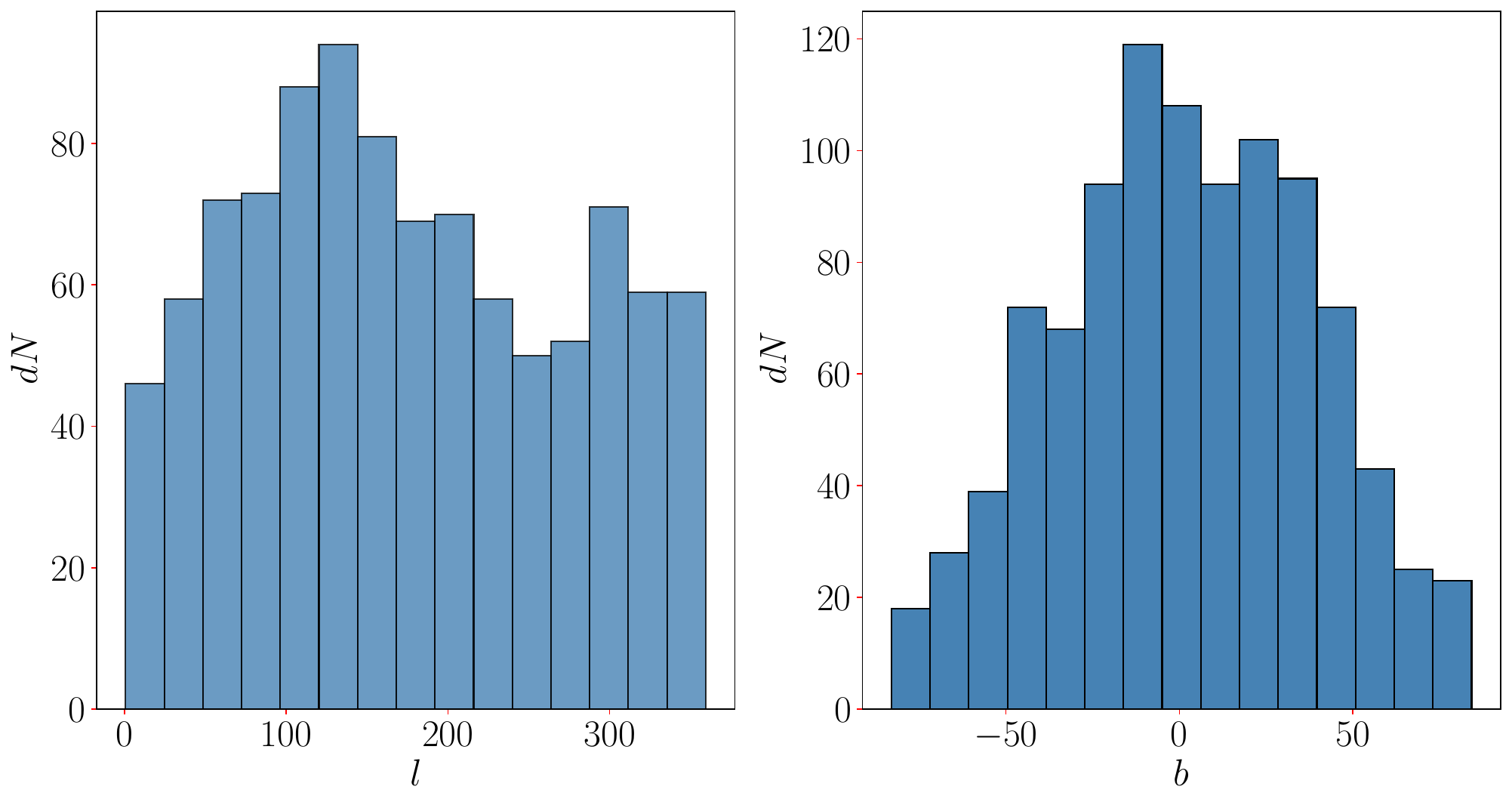}\includegraphics[width=0.5\linewidth]{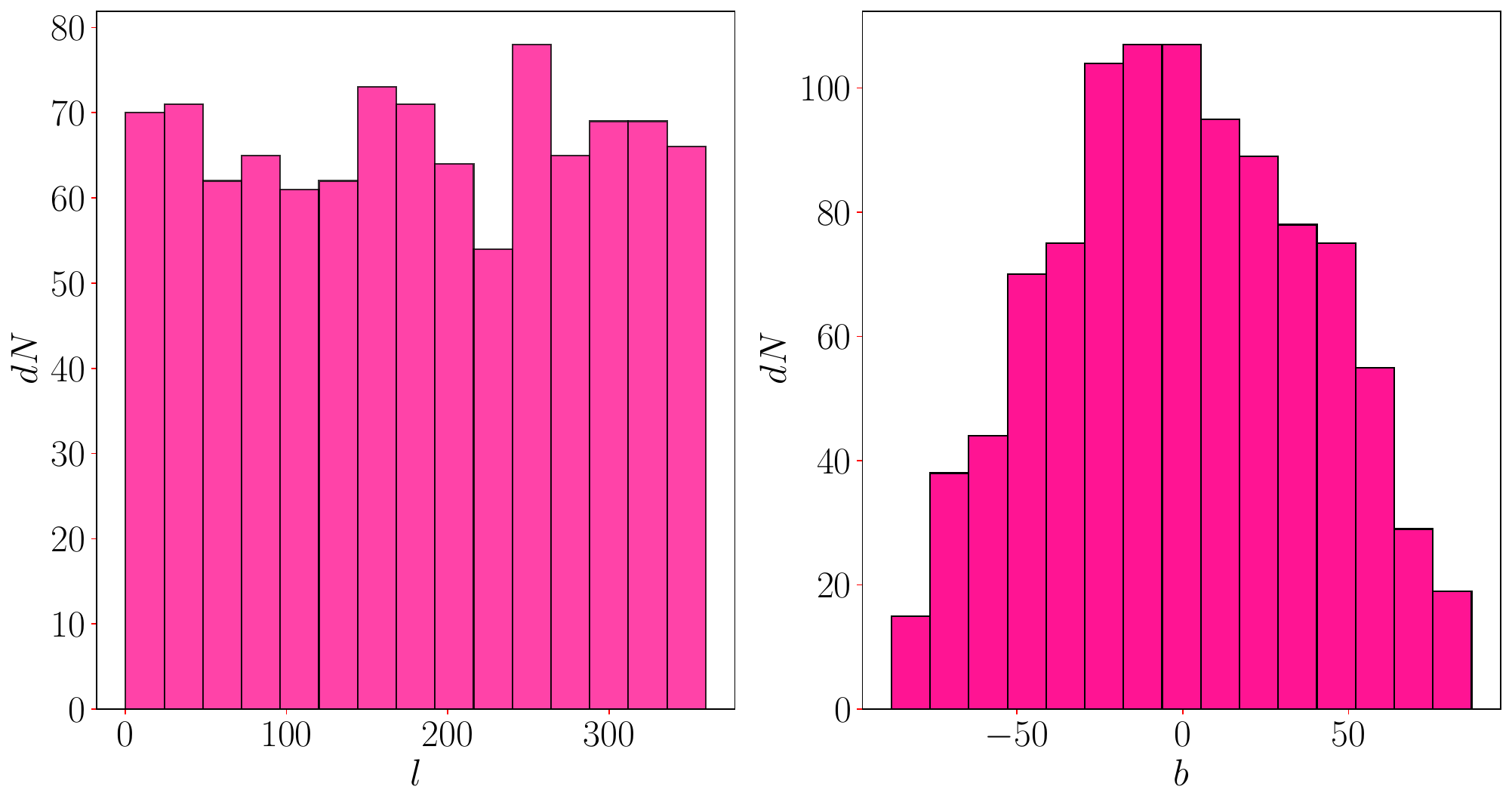}
\caption{Histogram of the distribution of galactic coordinates $l$ and $b$, in degrees, for the dipole directions in the generated random $H_0^{\text{Iso}}$-maps. 
The first and second panels analyze the random maps $H_{0}^{\text{Iso-48}}$-map (48 caps), 
while the third and fourth panels analyze the random maps 
$H_{0}^{\text{Iso-192}}$-map (192 caps). 
In both cases, our results confirm that the isotropic $H_0^{\text{Iso-48/192}}$-maps do not exhibit a preferred direction, 
as expected.}
\label{hist_coord_random}
\end{figure*}

Suppose that the number of SNe in the $i$th spherical cap is $n_i$, 
for $i = 1, \cdots, N$. 
These $n_i$ SNe were selected in the $i$th cap because their angular separation with respect to the center of the cap is less or equal $\gamma$ degrees. 
In this test, instead, for the $i$th cap we randomly choose $n_i$ SNe from any part of the sky, where $i = 1, \cdots, N$; then, we redo the best fit procedure in the $N$ HL diagrams obtaining an isotropic $H_0$-map, termed $H_0^{\text{Iso}}$-map. 
We repeat this process $1000$ times, obtaining 1000 $H_0^{\text{Iso}}$-maps. 
For illustrative purposes, in Figure~\ref{maps_random} we show two of these $H_0^{\text{Iso}}$-maps, both for $N=48$ and $192$ spherical caps, i.e., 
$H_0^{\text{Iso-48}}$-maps and $H_0^{\text{Iso-192}}$-maps, respectively. 
We also calculate the angular coordinates of the dipole components from these $H_0^{\text{Iso}}$-maps, and show them collectively as histograms in Figure~\ref{hist_coord_random}, again for both angular resolutions, that is, for $N=48$ and $N=192$ spherical caps. 

\begin{figure*}[t]
\begin{center}
\includegraphics[width=0.4\linewidth]{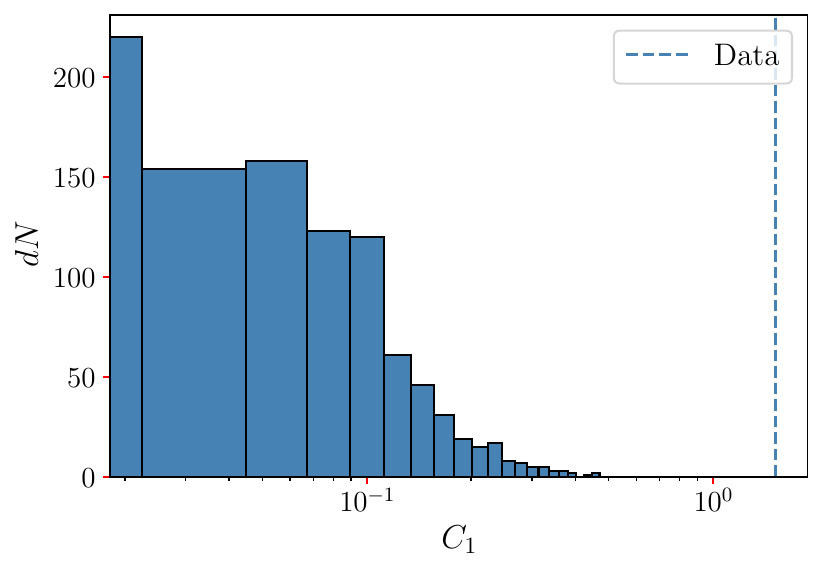}
\includegraphics[width=0.4\linewidth]{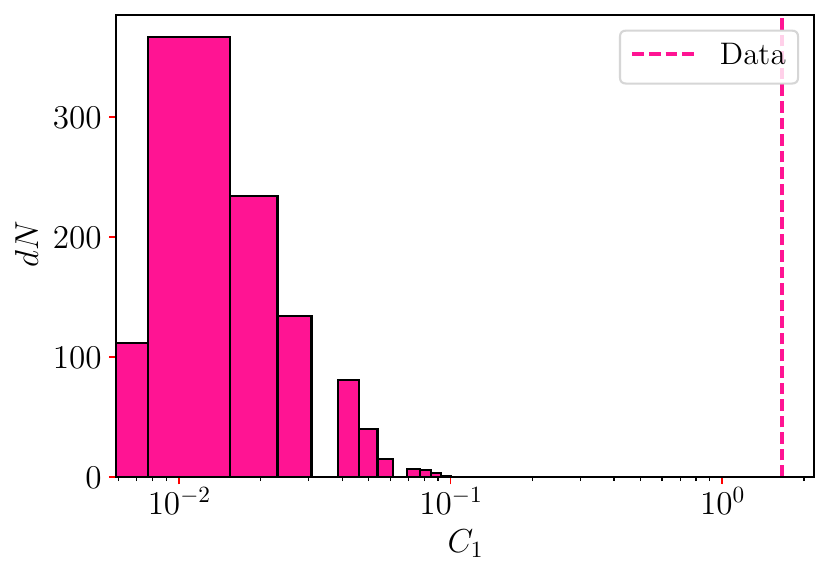}
\caption{Histograms of the dipole components, $C_1$, obtained from $1000$ randomized maps, plus the vertical dashed line draw for comparison 
from the $H_0$-map$^{48}$ ({\bf left panel}) and $H_0$-map$^{192}$ ({\bf right panel}), with $C_1=1.52$ and $C_1=1.66$ from the original maps, respectively. As observed, in both cases, our directional analysis show that 
the dipole components of the $H_0$-maps 
displayed in Figure~\ref{fig:H0maps}
are statistically significant, at more than $99.9\%$ confidence level.} 
\label{dipoles}
\end{center}
\end{figure*}

Furthermore, we calculate the corresponding dipole component of the $H_0^{\text{Iso}}$-maps in both angular resolutions, $\{ C_1^{\text{Iso-48/192}} \}$, 
useful information to calculate the statistical significance of the dipole components $C_1^{48}$ and $C_1^{192}$, from the $H_0$-map$^{48}$ and $H_0$-map$^{192}$ (shown in Figure~\ref{fig:H0maps}), respectively. 
The distribution of the dipole components from the $H_0^{\text{Iso-48}}$-maps, 
$\{ C_1^{\text{Iso-48}} \}$, and from the 
$H_0^{\text{Iso-192}}$-maps, 
$\{ C_1^{\text{Iso-192}} \}$, 
as well as the values $C_1^{48}=1.52$ 
and $C_1^{192}=1.66$, are displayed in Figure~\ref{dipoles}; from these distributions, we obtain the means and standard deviations: 
$C_1^{\text{Iso-48}} = 0.08$, $C_1^{\text{Iso-192}} = 0.02$
and $\sigma_{\text{Iso-48}}= 0.07, \, \sigma_{\text{Iso-192}}=0.02$. 

In Appendix~\ref{appendixA} we perform Monte Carlo simulations to calculate the uncertainties in $C_1^{48}$ and $C_1^{192}$, i.e., 
$\sigma_{C_1^{\text{Ran-48}}} = 0.67$ and $\sigma_{C_1^{\text{Ran-192}}} = 0.35$, respectively. 
With these data one can finally calculate the statistical significance of the dipole components found in our directional analysis; for instance, 
for the dipole case $C_1^{192}$ one computes the following: 
$|C_1^{192} - C_1^{\text{Iso-192}}| / \sqrt{\sigma_{C_1^{\text{Ran-192}}}^2 + \sigma_{\text{Iso-192}}^2}$, 
finding that this dipole is significant at more than $99.9 \%$ confidence level.

\subsection{Measuring the Hubble constant}

Although our directional approach is not intended to perform a $H_0$ measurement, 
even so, one can obtain this important information from the monopole components 
of our $H_0$-maps$^{48/192}$.\footnote{$H_0 = \sqrt{C_0 / 4\pi}$, where $C_0$ is the monopole term of the $H_0$-maps.} 
This allows us to calculate the Hubble constant at the effective distance 
$R^{192} = 102.83 \pm 10.24$~Mpc, that is, $z \simeq 0.025$, obtaining 
$H_0 = 70.39 \pm 1.40$ \text{~km~s$^{-1}$~Mpc$^{-1}$} using the $H_0$-map$^{192}$ and 
considering spherical caps with $\gamma=60^{\circ}$. 
For completeness, we also calculate for other angular resolutions and spherical cap sizes used in our scrutiny; 
the results are summarized in Table~\ref{results_H0}.

\begin{table*}[!ht]
\centering
\scalebox{1}{
\begin{tabular}{c|c}
\hline
$\gamma\,\backslash\,H_0$-map & $H_0$ [\text{~km~s$^{-1}$~Mpc$^{-1}$}] \\
\hline
$\gamma=60^{\circ}\,\backslash\,H_0$-map$^{48}$  & $70.35 \pm 1.50$  \\
$\gamma=60^{\circ}\,\backslash\,H_0$-map$^{192}$ & $70.39 \pm 1.40$  \\
\hline
$\gamma=65^{\circ}\,\backslash\,H_0$-map$^{48}$ & $70.42 \pm 1.45$  \\
$\gamma=65^{\circ}\,\backslash\,H_0$-map$^{192}$ & $70.39 \pm 1.27$  \\
\hline
$\gamma=70^{\circ}\,\backslash\,H_0$-map$^{48}$ & $70.43 \pm 1.38$  \\
$\gamma=70^{\circ}\,\backslash\,H_0$-map$^{192}$ & $70.41 \pm 1.18$  \\
\hline 
\end{tabular}
}
\caption{$H_0$ measurements for different spherical cap sizes.}

\label{results_H0}
\end{table*}

In Appendix~\ref{ap:hubble-tension} we discuss the influence that overdensity and underdensity matter structures can produce on $H_0$ measurements.


\section{Conclusions}
\label{sec:final}

Cosmic matter structures grow over time due to gravitational interaction. 
This explains why our cosmic neighborhood, $z \simeq 0$, exhibits amazingly 
large underdense and overdense matter structures \citep{rubin1951,
gerard1953,Courtois13,Hoffman2017,Tully19,Colgain2019,deCarvalho20,Avila21,Avila22,Marques24}. 
In this scenario, it was recently proposed that the DR and the Shapley supercluster structures dominate the galaxy velocity field in the Local Universe, acting like a dipolar gravitational system. 
This scenario deserves investigation because 
in the Local Universe the effect of peculiar motions caused by the distribution of matter competes with the expansion of the universe making the analysis of the HL diagram difficult, affecting the measurement of the Hubble constant. 

Motivated by this proposal, we investigated the subsample of 501 SNe Ia of the Pantheon+ catalog in the Local Universe, with redshifts $0.015 \leq z \leq 0.06$, aiming to measure both the direction and magnitude of the bulk flow velocity. 
Our directional analysis scan the celestial sphere considering two angular resolutions: with $N=48$ and $N=192$ spherical caps, although our conclusions are drawn from the best angular resolution case, i.e., for $N=192$ caps. 
We found a statistically significant dipole variation of the Hubble constant, at more than 99.9\% confidence level, with $H_0$ values in the $+/-$ dipole directions: $H_{0}^{+/192} = 72.31 \pm 1.34$ ~km~s$^{-1}$~Mpc$^{-1}$ and $H_{0}^{-/192} = 69.74 \pm 1.30$ ~km~s$^{-1}$~Mpc$^{-1}$, for the $H_0^{192}$-map. 
The Table~\ref{results_table}, and Figures~\ref{fig:H0maps} and~\ref{bulkflow}, summarize our results. 

Our studies confirm that matter structures in the Local Universe are indeed following a dipolar bulk flow motion toward $(l,b) = (326.^\circ1 \pm 11.^\circ2,27.^\circ8 \pm 11.^\circ2)$, 
a direction close to the Shapley supercluster ($l_{\scalebox{0.7}{Shapley}}, b_{\scalebox{0.7}{Shapley}}) = (311.53^\circ, 32.31^\circ$), with the velocity of $132.14 \pm 109.30$ ~km~s$^{-1}$ at the effective distance $102.83 \pm 10.24$ Mpc = $71.98 \pm 7.17$ Mpc~$h^{-1}$, using $h=0.7$ to illustrate the measurements' comparison shown in Figure~\ref{bulkflow}. 

Not less interesting, the antipodal direction of the dipolar bulk flow found points close to the DR structure, as illustrated in Figure~\ref{resultsky}. 

As shown in Figure~\ref{bulkflow}, 
recent studies of the CosmicFlows-4 data \cite[CF4,][]{Tully23} 
at scales greater than $100$ Mpc~$h^{-1}$ have reported bulk flow velocities that are in great tension with the velocities expected in the standard cosmological model, like the measurements of \cite{Watkins23} who found  $395 \pm 29$ ~km~s$^{-1}$ at 150 Mpc~$h^{-1}$ and 
$427 \pm 37$ km~s$^{-1}$ at 200 Mpc~$h^{-1}$, and the measurement of~\cite{Whitford23} who found $428 \pm 108$ ~km~s$^{-1}$ at 173 Mpc~$h^{-1}$, values with very low probability 
of occurrence: $0.015\%$, $0.00015\%$, and $0.11\%$, respectively 
(these data are also displayed in Figure~\ref{bulkflow}). 
As suggested by \cite{Watkins23}, these results may be an indication 
that the matter rest frame of the universe is not that determined from the dipole in the CMB 
radiation~\cite[see, e.g.,][]{Migkas21}. 

However, several analyses performed at moderate scales $\lesssim 100$ Mpc~$h^{-1}$ estimate the bulk flow velocities finding that they are statistically consistent with the $\Lambda$CDM prediction, as well as 
the measurement of~\cite{Whitford23} at $49$ Mpc~$h^{-1}$ analysing the CF4 dataset, 
and also our measurement at the effective distance of $72$~Mpc~$h^{-1}$.

Concluding, our directional analysis shows that the bulk flow velocity field in the Local Universe is well explained by the gravitational dipole system {\em Shapley-DR}~\citep{Hoffman2017}. 
In addition, we are confident that our results are robust, and the methodology adopted is unbiased due to the various consistency tests done, which we present in Appendix.

\end{multicols}


 \section*{Acknowledgments}
M.L., C.F., and A.B. acknowledge to CAPES and CNPq for their corresponding fellowships. 
F.A. thanks CNPq and FAPERJ, Processo SEI 260003/014913/2023, for financial support. 

\appendix

\section{Consistency test for other redshift intervals}\label{ap:redshift-intervals}

\begin{figure}[!ht]
\centering
\includegraphics[width=0.45\linewidth]{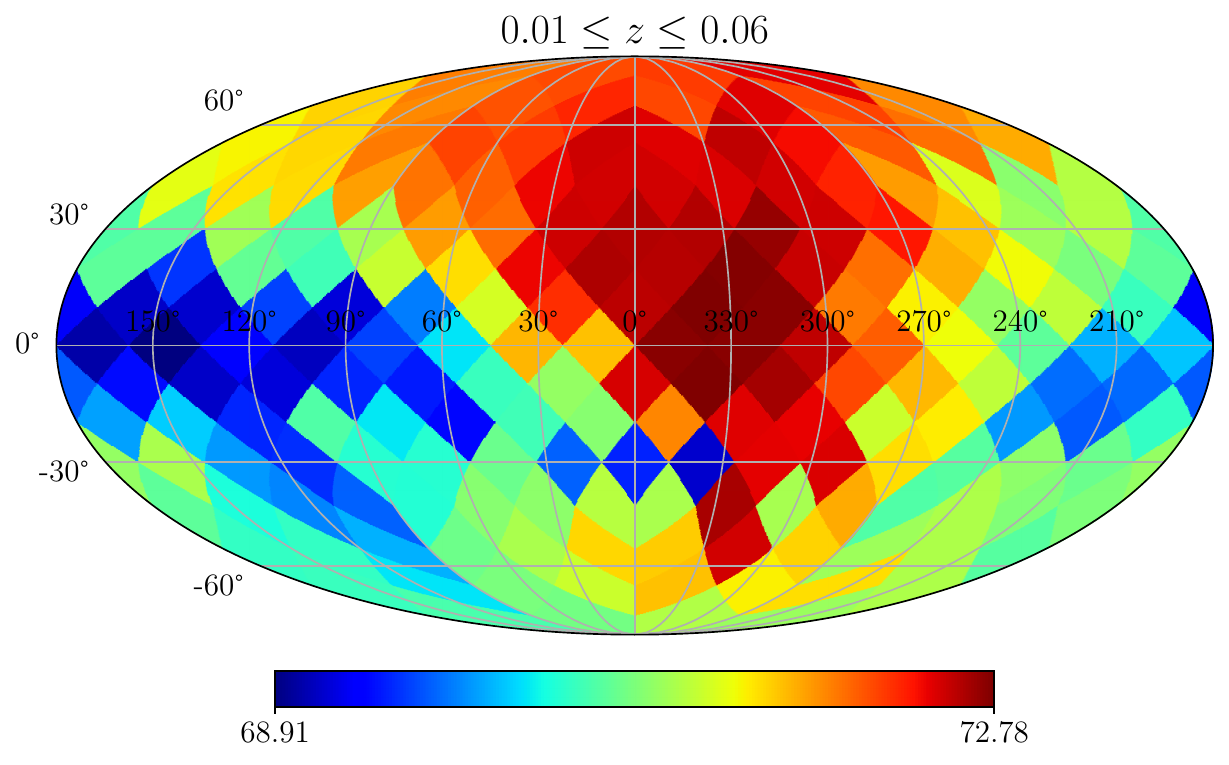}
\includegraphics[width=0.45\linewidth]{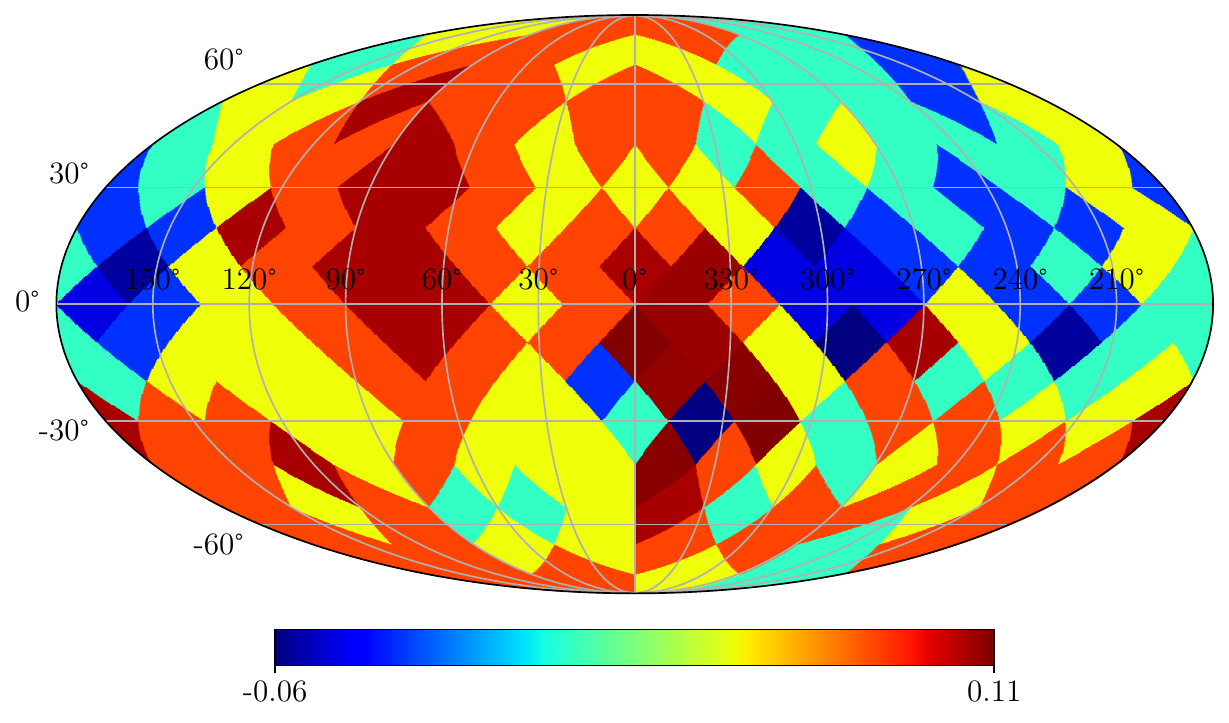} \\
\includegraphics[width=0.45\linewidth]{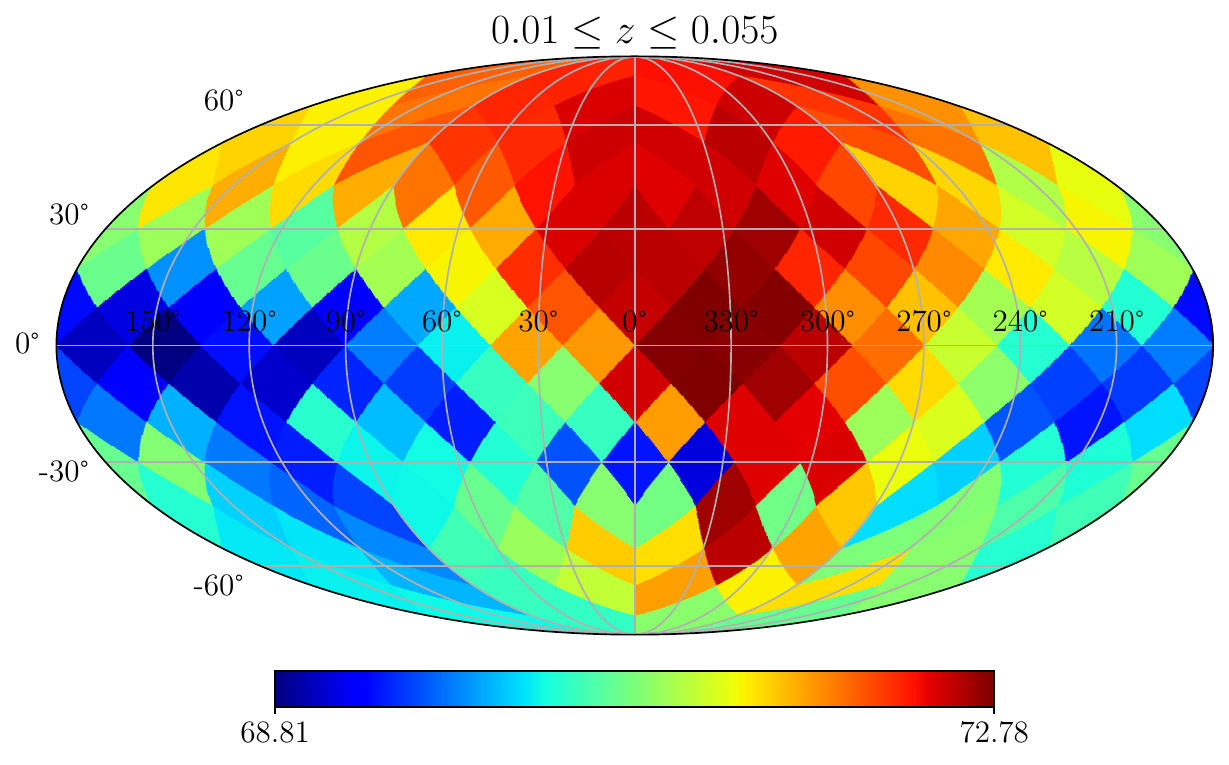}
\includegraphics[width=0.45\linewidth]{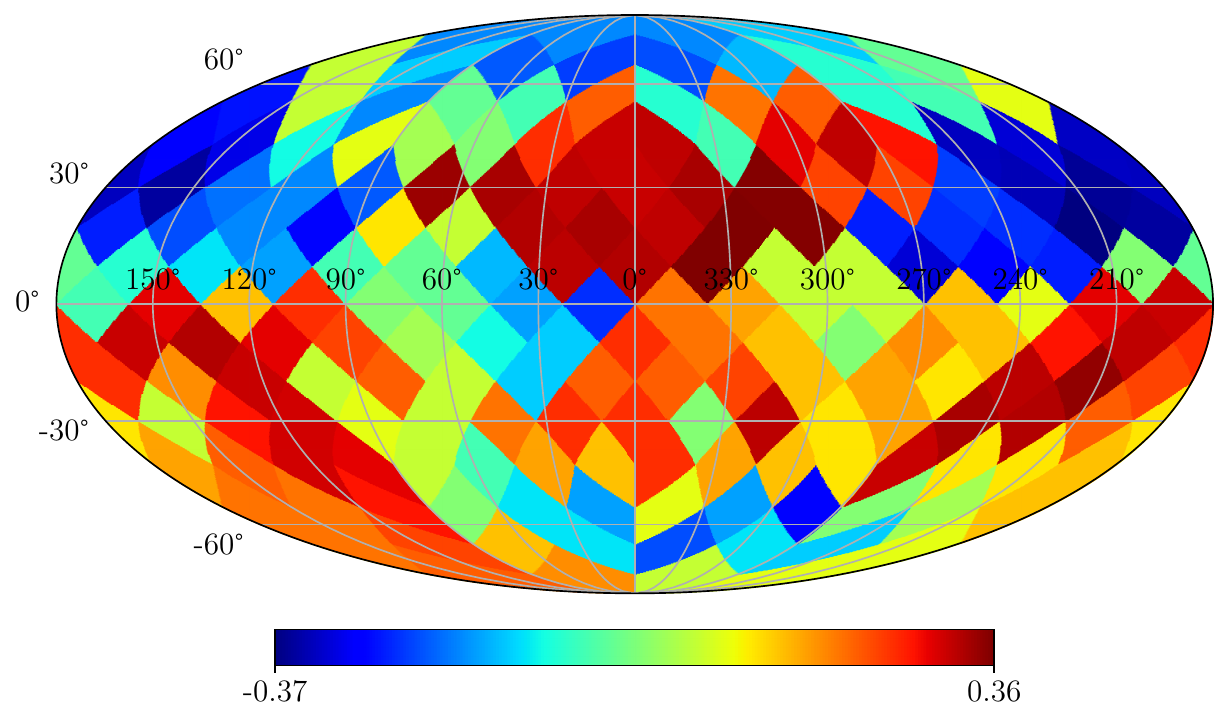} \\
\includegraphics[width=0.45\linewidth]{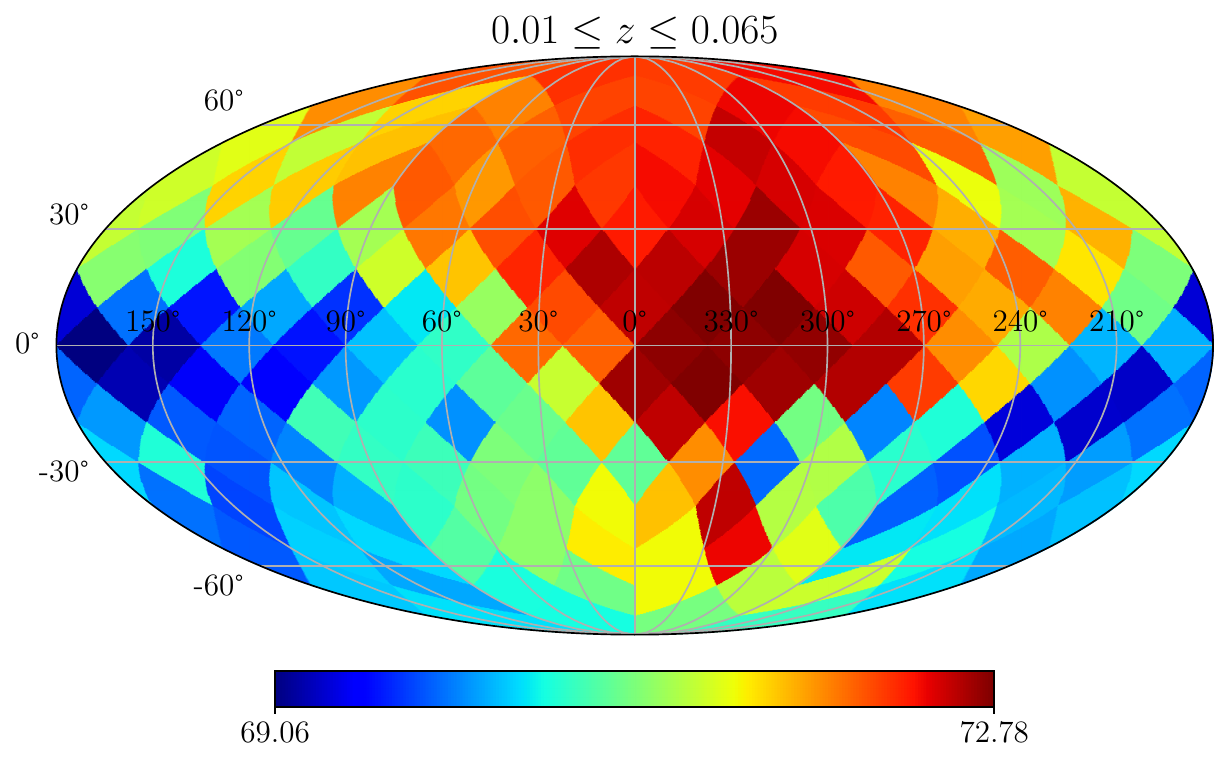}
\includegraphics[width=0.45\linewidth]{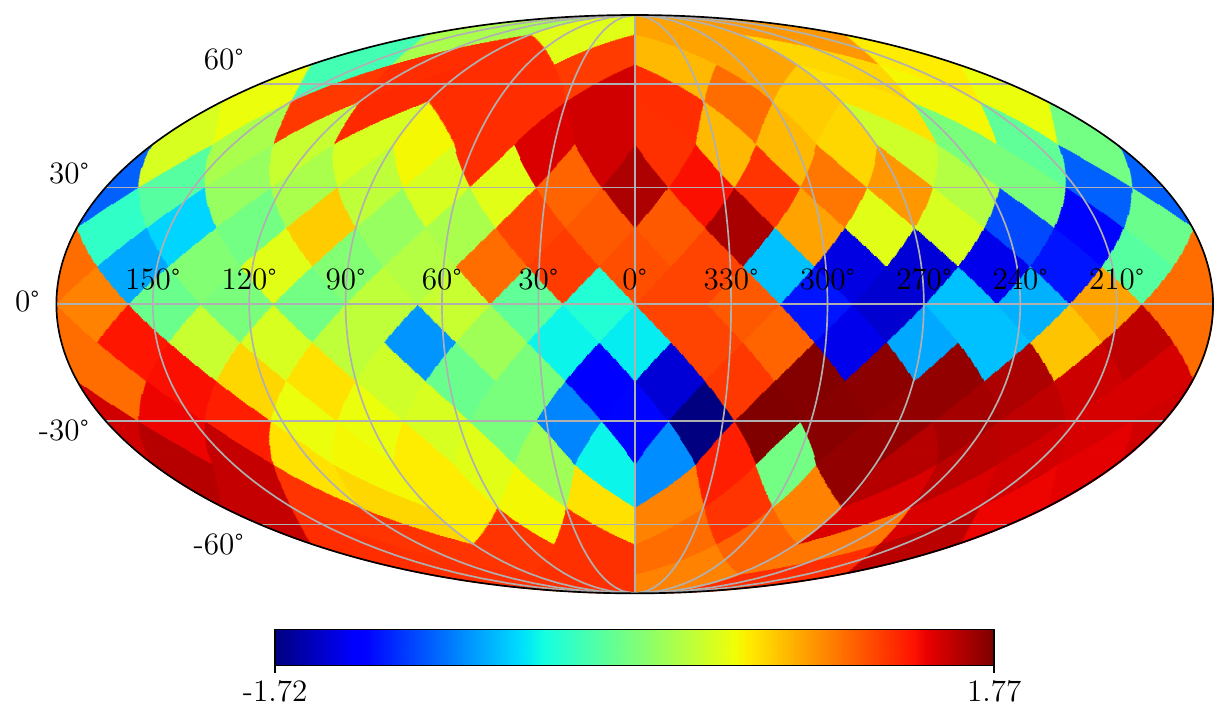} \\
\includegraphics[width=0.45\linewidth]{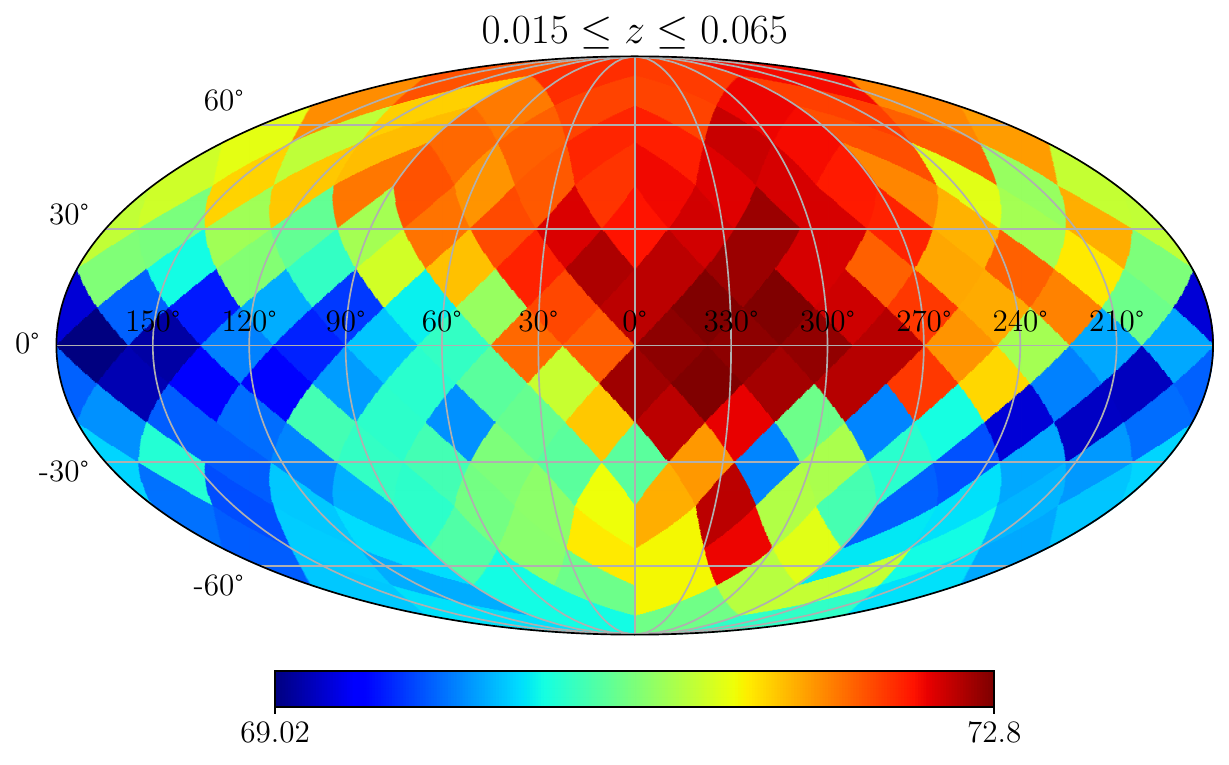}
\includegraphics[width=0.45\linewidth]{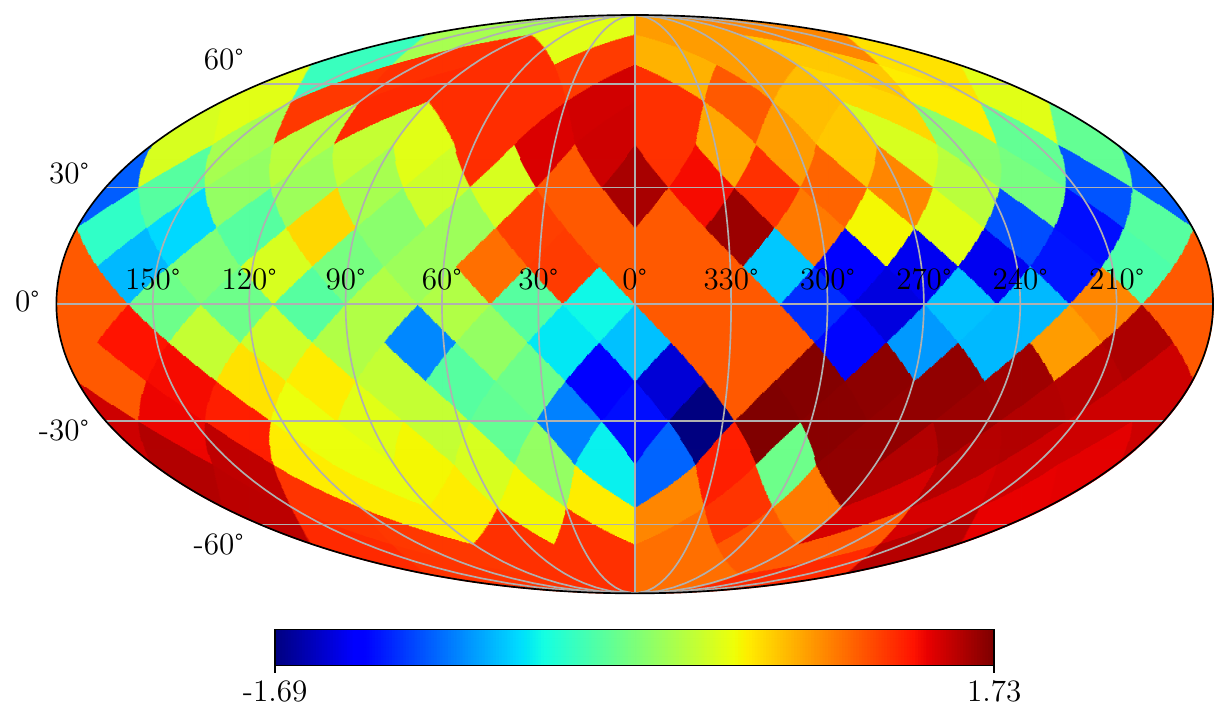} \\
\caption{\textbf{Left column:} the $H_0$-maps$^{192}$ for different redshift intervals. \textbf{Right column:} the maps of the difference between the original $H_0$-map$^{192}$ (see Figure \ref{fig:H0maps}) and the corresponding map to the left.}
\label{fig:redshift-intervals}
\end{figure}

We have investigated the consistency of our results studying the SNe present in other redshift intervals, close to the original interval analyzed in this work. 
Thus, we performed our directional analysis for four SNe samples with redshift intervals as follows: 
$z \in [0.01,0.06]$, $z \in [0.01,0.055]$, 
$z \in [0.01,0.065]$, and $z \in [0.015,0.065]$; 
the number of SNe Ia in these intervals was 565, 552, 576, and 512, respectively. 
The $H_0$-maps$^{192}$ obtained in each case are displayed in the 
left column of Figure~\ref{fig:redshift-intervals}. 
To illustrate how different these maps are with respect to the 
original $H_0$-map$^{192}$, analyzed for the SNe with 
$z \in [0.015,0.06]$ and displayed in the left panel, second row, of Figure~\ref{fig:H0maps}, we calculate the map difference 
$D^i \equiv$ [$H_0$-map$^{192/\text{orig}}$ - $H_0$-map$^{192/i}$], where the index $i$ refers to each one of the four samples mentioned above. 
The maps $\{ D^i \}$, displayed in the right column of Figure~\ref{fig:redshift-intervals}, 
show that the differences with respect to $H_0$-map$^{192/\text{orig}}$ are very tiny for the first two cases, and differences $\lesssim 2 \%$ for the last two cases, 
implying that our results regarding the interval analyzed in this work, $z \in [0.015,0.06]$, are robust.

\section{Uncertainty Due to the Dipole \texorpdfstring{$\delta H_0$}{}: Monte Carlo Simulations}\label{appendixA}

The dipolar nature of $\delta H_{0}(l,b)$ contributes to the uncertainty in the measurement of $H_{0}^{b\text{-}\!f}$ (see Section~\ref{uncertainties}). 
To calculate this uncertainty, $\sigma_{dp}$, due to the directional dependence observed in $H_0^{b\text{-}\!f}\!(l,b)$, we perform Monte Carlo simulations. 

We first generate $1000$ random samples of SNe, where their angular positions $(l, b)$ were preserved, but their luminosity distances were estimated randomly from a Gaussian distribution, considering, as the mean and the standard deviation of such distribution, the luminosity distance of each SN $D_L$, 
and its uncertainty $\sigma_{D_L}$, respectively. 
We then performed the same process of directional analysis considering both angular resolutions, that is, with $N=48$ and $N=192$ caps. 
Then, performing the $H_0$ best fit procedure in the HL diagrams, we obtain $1000$ random $H_0$-maps$^{\text{Ran-48}}$ and $1000$ random 
$H_0$-maps$^{\text{Ran-192}}$.
\footnote{Not to be confused with the isotropic 
$H_0^{\text{Iso-48/192}}$-maps, see Section~\ref{isotropic-maps}.} 

\begin{figure*}
\begin{center}
\includegraphics[width=0.33\linewidth]{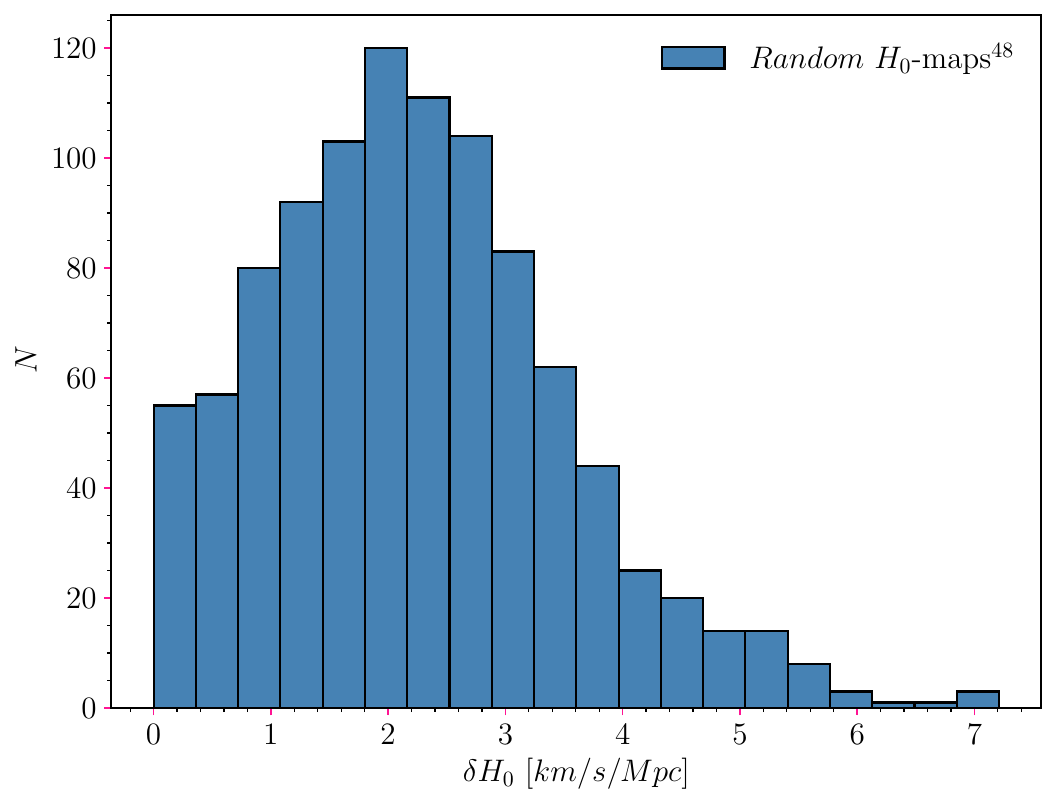}\includegraphics[width=0.33\linewidth]{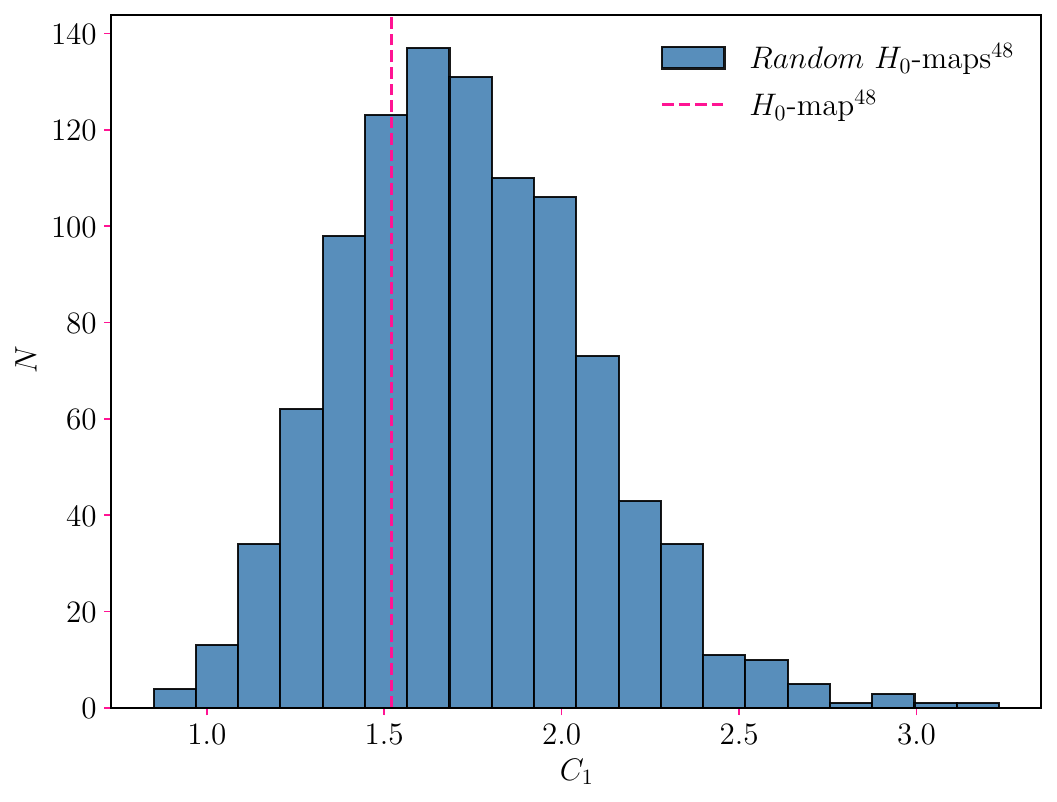} \\ \includegraphics[width=0.33\linewidth]{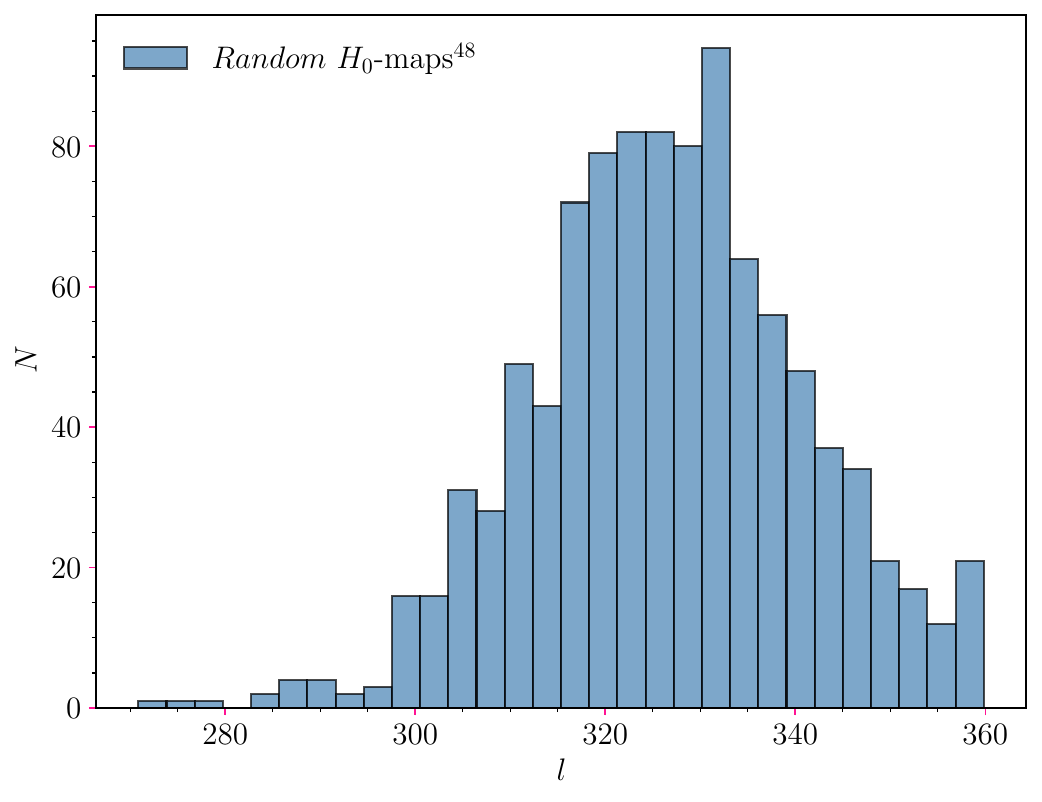} \includegraphics[width=0.33\linewidth]{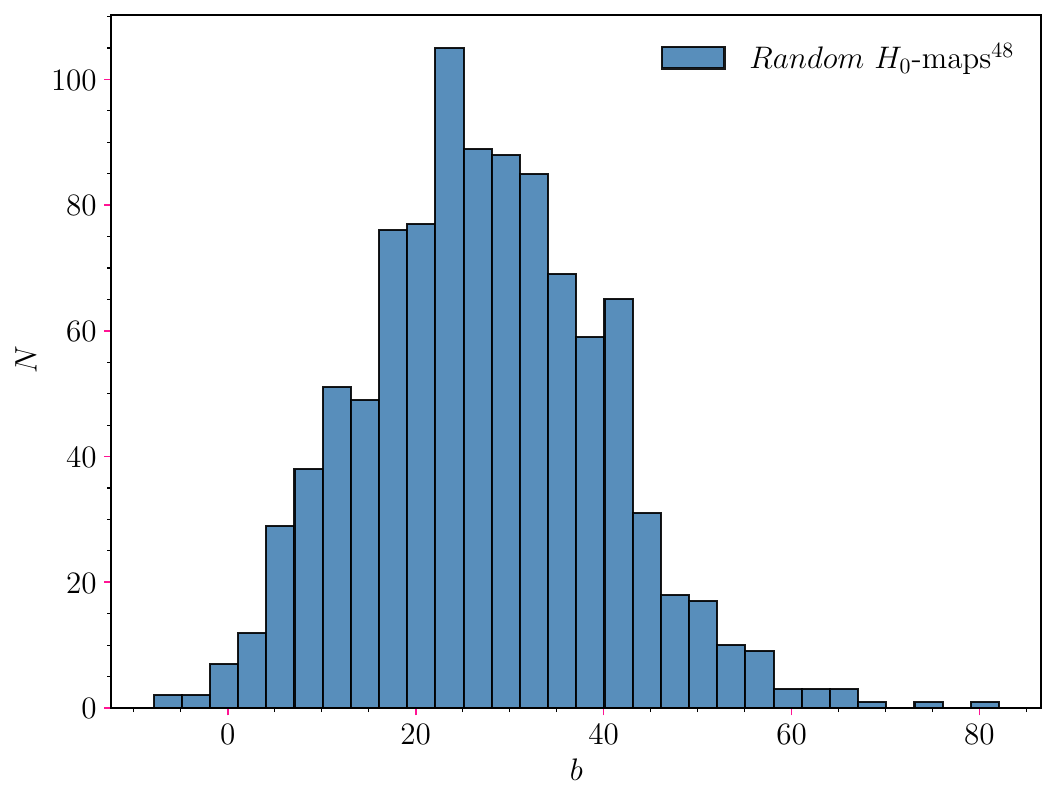}\\
\includegraphics[width=0.33\linewidth]{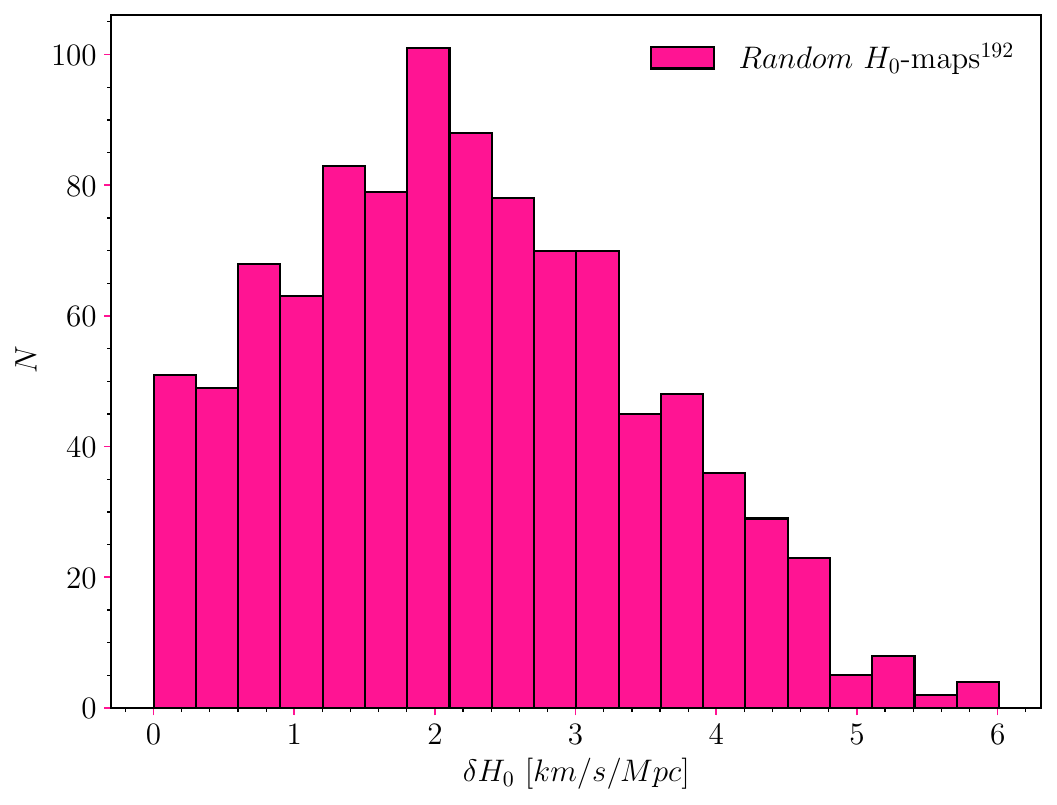}\includegraphics[width=0.33\linewidth]{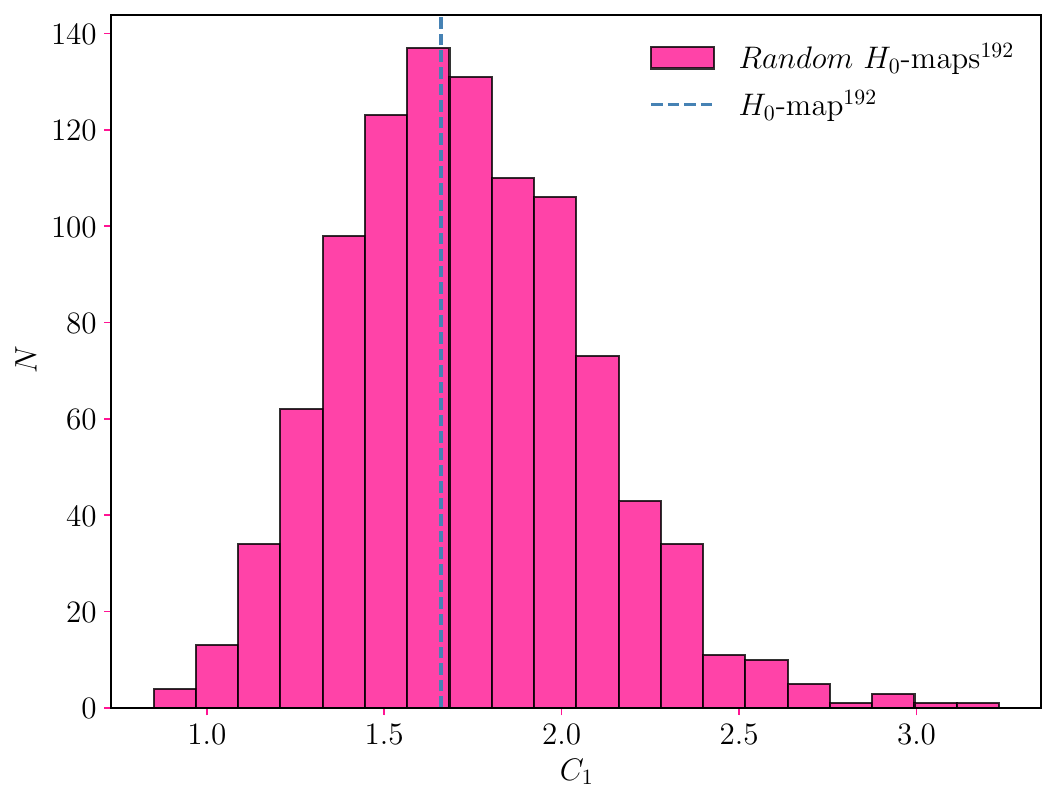} \\ \includegraphics[width=0.33\linewidth]{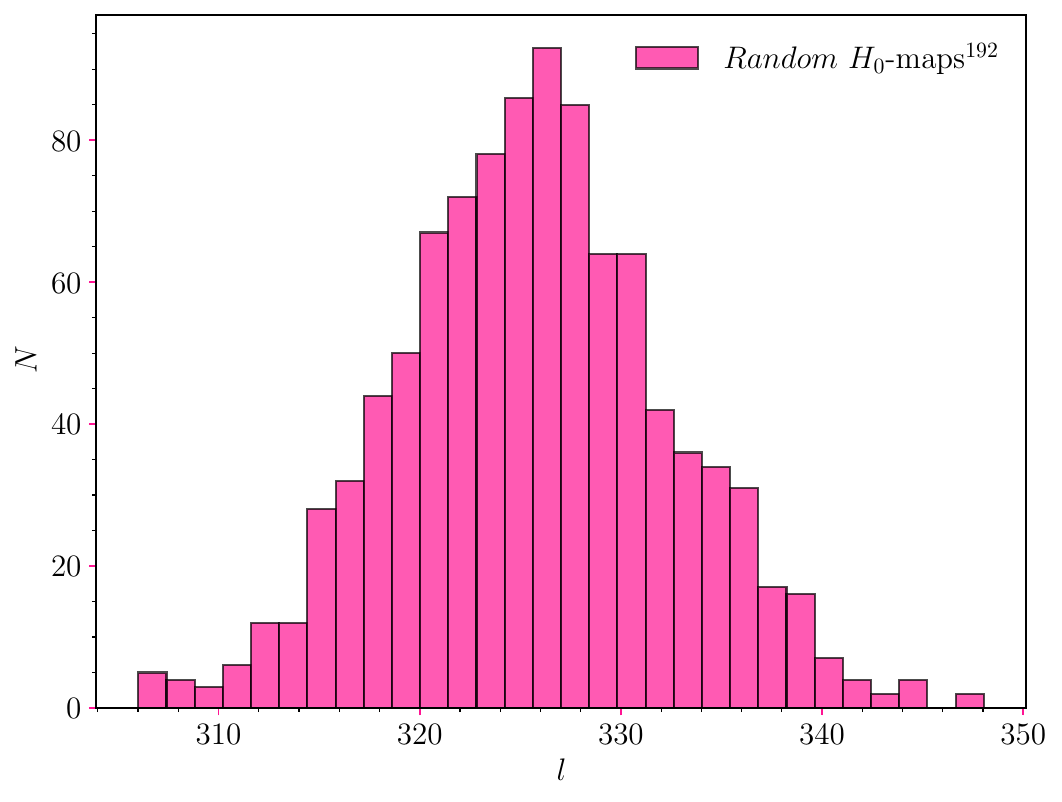} \includegraphics[width=0.33\linewidth]{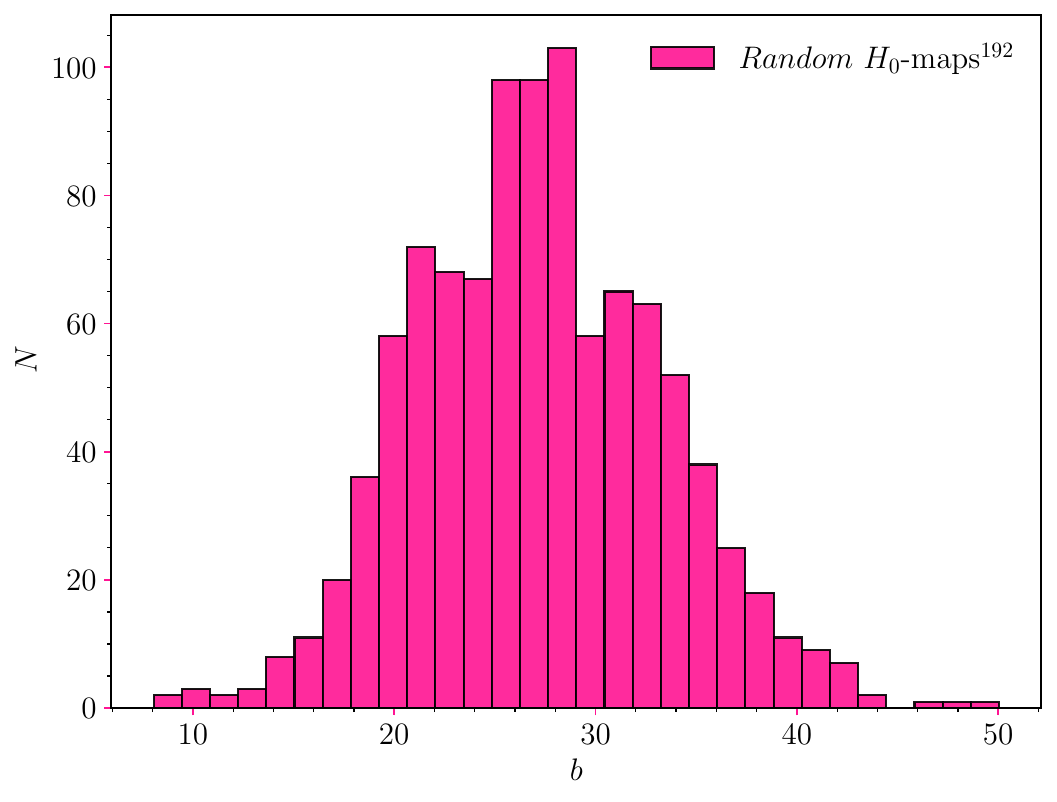}
\end{center}
\caption{ 
These plots show the results of our Monte Carlo simulations performed to produce 
$H_0$-maps$^{\text{Ran-48/192}}$ from which one can estimate the uncertainties of quantities used in our approach. 
In the first row we display the amplitudes $\{ \delta H_0 \}^{\text{Ran-48}}$ (left panel) and the dipole component $\{ C_1 \}^{\text{Ran-48}}$ (right panel) calculated from random $H_0$-maps$^{\text{Ran-48}}$, where their standard deviations provide 
the uncertainties 
$\sigma_{dp}^{48} = 1.26$ ~km~s$^{-1}$~Mpc$^{-1}$, and 
$\sigma_{C_1^{\text{Ran-48}}} = 0.67$, respectively (see Appendix~\ref{appendixA} for details). 
The third row show similar calculations, but for the $N=192$ case, namely, 
$\sigma_{dp}^{192} = 1.24$ ~km~s$^{-1}$~Mpc$^{-1}$, and 
$\sigma_{C_1^{\text{Ran-192}}} = 0.35$. 
The second and fourth rows provide the distributions of the dipole directions of the $H_0$-maps$^{\text{Ran-48/192}}$, showing mean and standard deviations as expected due to the procedure used in these Monte Carlo simulations, that is, 
$(l, b)^{\text{Ran-48}}  = 
(326.^{\circ}58 \pm 14.^{\circ}33, 27.^{\circ}33 \pm 12.^{\circ}76)$ and 
$(l, b)^{\text{Ran-192}} = 
(325.^{\circ}68 \pm 6.^{\circ}76, 27.^{\circ}17 \pm 6.^{\circ}1)$. 
The vertical dashed lines in the $\{ C_1 \}^{\text{Ran-48/192}}$ plots 
(right panels in the first and third row) represent the dipole values $C_1^{48}=1.52$ and $C_1^{192}=1.66$ of the $H_0$-maps displayed in Figure~\ref{fig:H0maps} (see Section~\ref{isotropic-maps} and Figure~\ref{dipoles}).
}
\label{figD1}
\end{figure*}

For each one of these maps, we calculate the amplitudes $\{ \delta H_0 \}^{\text{Ran-48/192}}$, the dipole components $\{ C_1 \}^{\text{Ran-48/192}}$, and the dipole directions $\{ (l,b) \}^{\text{Ran-48/192}}$.
The corresponding histograms of the distributions obtained are shown in Figure~\ref{figD1} (the first and second rows correspond to $N=48$, and the third and fourth rows correspond to $N=192$).
The standard deviation of the distributions 
$\{ \delta H_0 \}^{\text{Ran-48/192}}$ provides a measure of the 
uncertainty due to the dipole nature of the $H_0$-map, $\delta H_0$, where we obtain $\sigma_{dp}^{48} = 1.26$ ~km~s$^{-1}$~Mpc$^{-1}$ and $\sigma_{dp}^{192} = 1.24$ ~km~s$^{-1}$~Mpc$^{-1}$, respectively.

Another quantity that we analyze in Figure~\ref{figD1} are the set of dipole components $\{ C_1 \}^{\text{Ran-48/192}}$ of the 
$H_0^{\text{Ran-48/192}}$-maps, where their standard deviations provide a measurement of the uncertainties in the dipole components $C_1^{48}$ and $C_1^{192}$ of our $H_0$-maps (shown in Figure~\ref{fig:H0maps}), 
obtaining $\sigma_{C_1^{\text{Ran-48}}} = 0.67$ and $\sigma_{C_1^{\text{Ran-192}}} = 0.35$, respectively. 
These quantities are used to estimate the statistical significance of the dipole nature of our results (see Section~\ref{isotropic-maps}). 

\section{Studying the correlation between the number of SNe 
and the best fit \texorpdfstring{$H_0$}{}}\label{pearson}

An interesting question regards the possible bias of the number of SNe 
observed in different directions could have an impact in the best fit analysis of the HL diagram to obtain $H_0$. 

\begin{figure*}[!ht]
\centering
\includegraphics[width=0.5\linewidth]{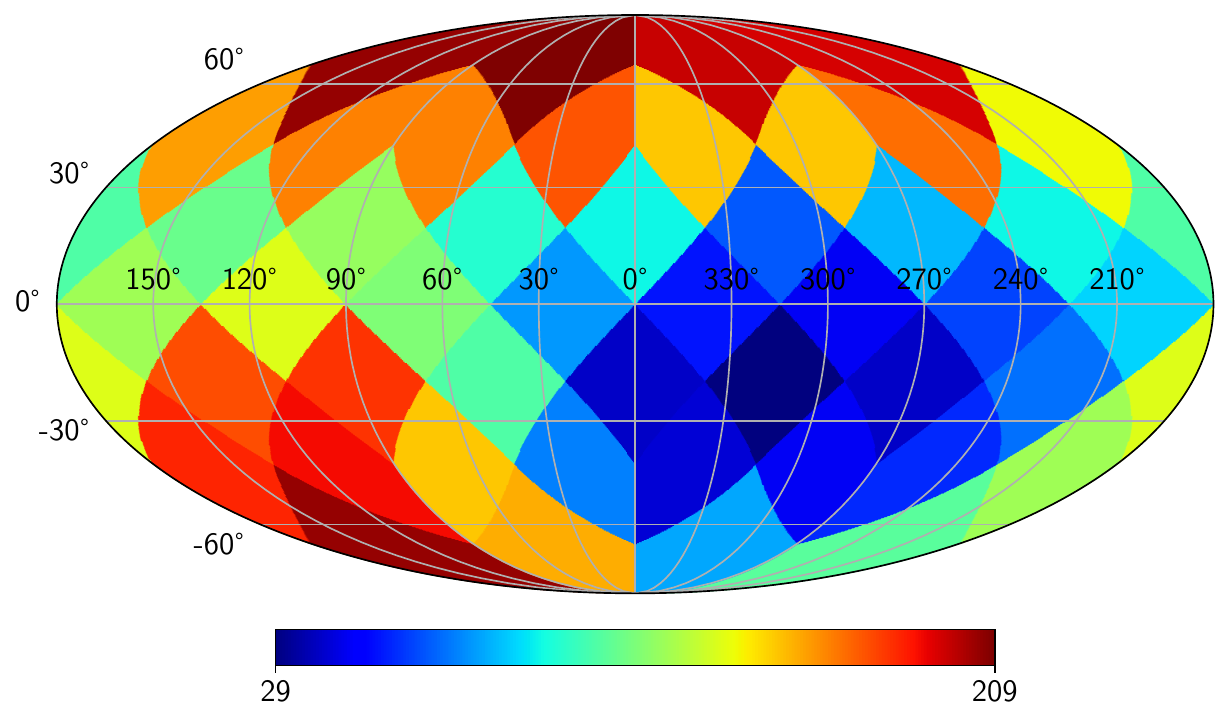}\includegraphics[width=0.5\linewidth]{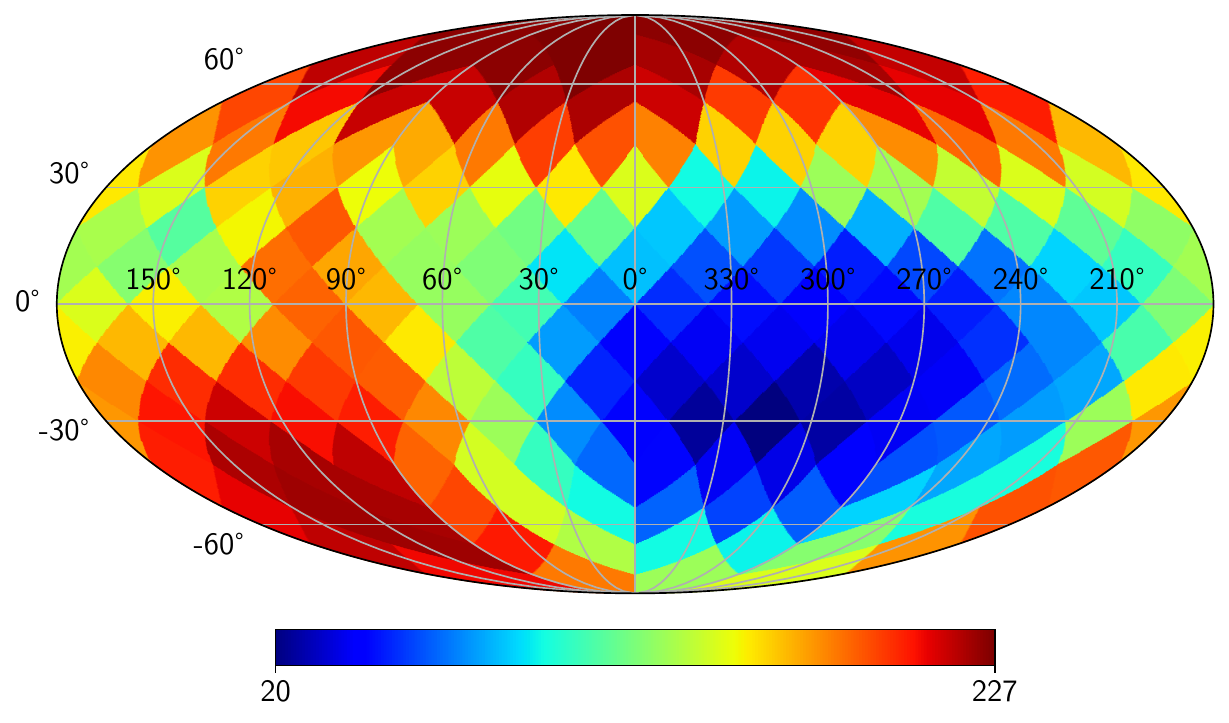}
\caption{Number of SNe maps, for 48 (left) and 192 (right) spherical caps analyses. 
These maps show the number of SNe analyzed in each spherical cap, 
as in the directional analysis procedure, by assigning a color to the corresponding number of SNe inside the $60^{\circ}$ cap. 
These maps are useful to study the possible correlation with the maps shown in Figure~\ref{fig:H0maps}; our results, discussed in Appendix~\ref{pearson}, confirm that the correlation is very weak.
}
\label{number-maps}
\end{figure*}

For this, we perform a correlation analysis of the $H_0$-map with the number-of-SNe-map 
(that is, the map where the color in each pixel 
represents the number of SNe used to construct our 
$H_0$-map, see Figure~\ref{number-maps}),  using the linear Pearson correlation coefficient 
$\cal{P}$. 
Our analyses show (the bars mean absolute value): 
$|{\cal P}| = 0.234$, for the case with $N=48$ spherical caps, and 
$|{\cal P}| = 0.198$, for the case with $N=192$ spherical caps. 
According to the literature, for values of the Pearson coefficient in the interval, 
$|{\cal P}| \in [0.0, 0.199]$ means that the correlation between the pairs of maps 
with the same angular resolution is {\em very weak}.

Notice that, for consistency, our analyses scan the celestial sphere considering two angular resolutions: with $N=48$ and with $N=192$ spherical caps. 
However, our conclusions are drawn from the best angular resolution case, i.e., scanning the celestial sphere with $N=192$ spherical caps. 
Due to this, we are led to conclude that the number of SNe observed in the 192 spherical caps 
has a negligible impact in the --statistically significant--dipolar behavior that our directional analysis found in the $H_0$-map. 


\section{Robustness for Other Spherical Cap Sizes: 
\texorpdfstring{$\gamma = 65^{\circ},\, 70^{\circ}$}{}}\label{othercaps}

In this appendix, we perform robustness tests to verify our directional analyses results considering spherical caps with radius different from $\gamma=60^{\circ}$, the case considered along this work, namely $\gamma=65^{\circ}$ and $\gamma=70^{\circ}$. 
Moreover, we make these tests for both angular resolutions, with 48 and 192 spherical caps. 

\begin{figure*}
\begin{center}
\includegraphics[width=0.45\textwidth]{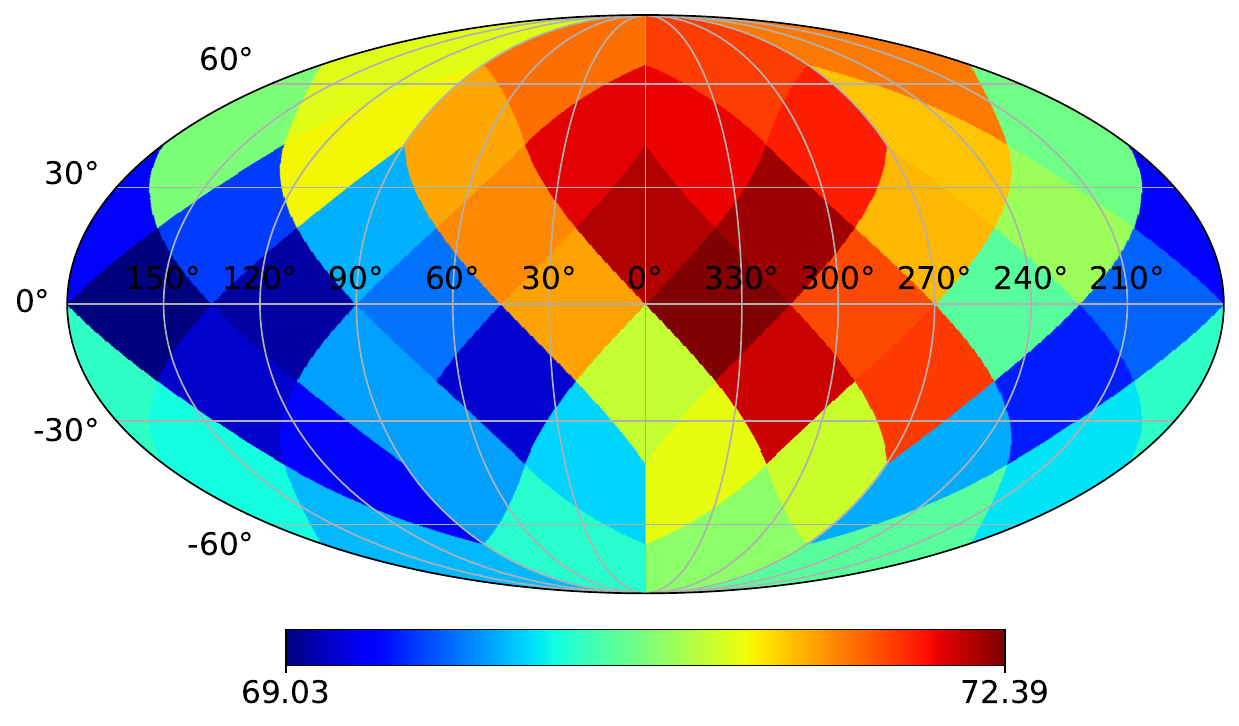}
\includegraphics[width=0.45\textwidth]{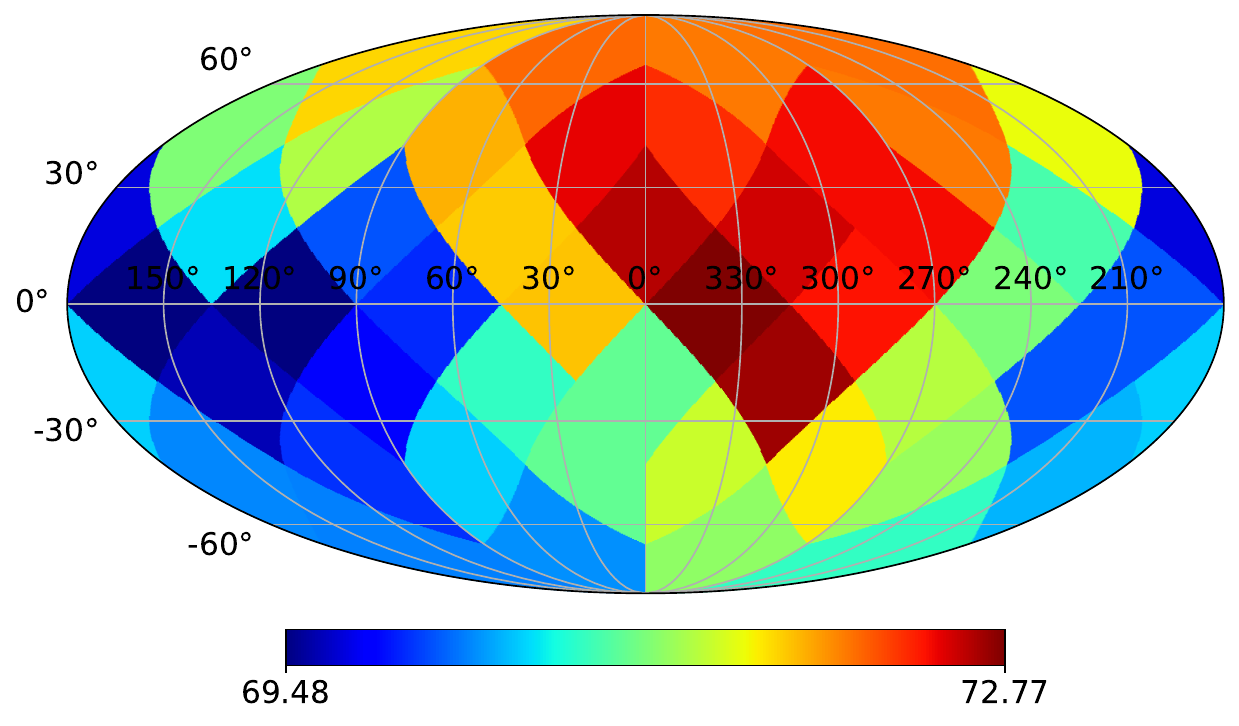} \\
\includegraphics[width=0.45\textwidth]{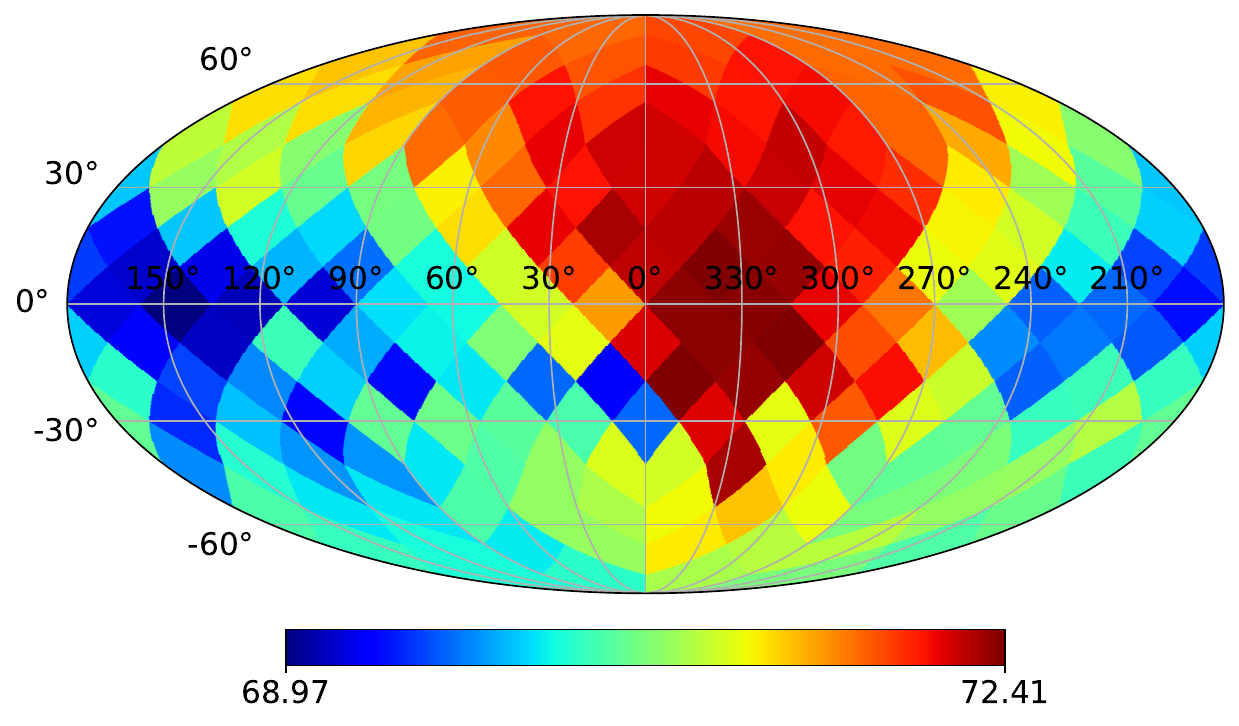}
\includegraphics[width=0.45\textwidth]{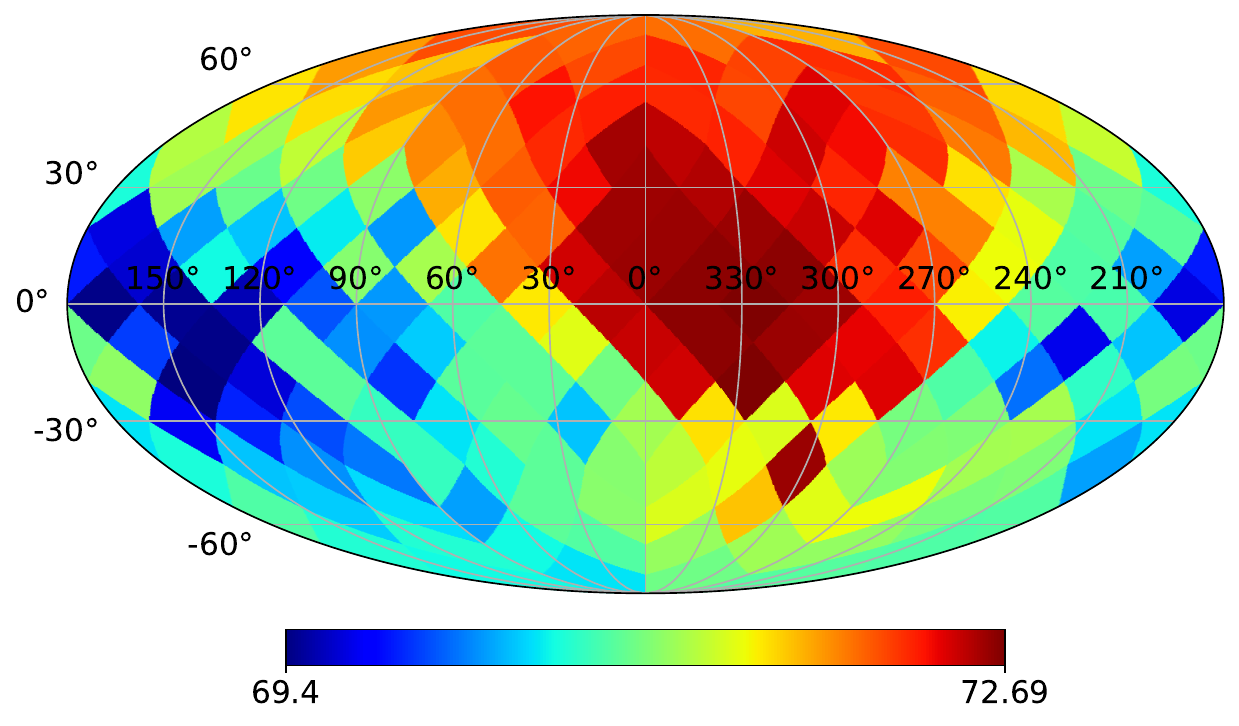}
\caption{
Robustness tests to confirm our directional analysis results considering different 
spherical cap sizes, namely: $\gamma=65^{\circ}$ and $\gamma=70^{\circ}$, tests done 
for both angular resolutions, with 48 and 192 spherical caps. 
The plots at the top correspond to the cases with 48 caps for $\gamma=65^{\circ}$ (left) 
and $\gamma=70^{\circ}$ (right). 
The plots at the bottom correspond to the cases with 192 caps for $\gamma=65^{\circ}$ (left) 
and $\gamma=70^{\circ}$ (right). 
As observed, comparing the dipole directions of these $H_0$-maps, displayed in 
Table~\ref{results_table2}, with those in Table~\ref{results_table} both results are 
in excellent agreement. 
}
\label{H0-maps2}
\end{center}
\end{figure*}

Observing Table~\ref{results_table2}, and the $H_0$-maps displayed in 
Figure~\ref{H0-maps2}, we obtain an excellent agreement with 
the results found along the text considering spherical caps with $\gamma=60^{\circ}$. 
Therefore, we conclude that our results, summarized in Table~\ref{results_table}, are 
robust considering directional analysis done with spherical caps of different sizes used to scan 
the celestial sphere. 
\begin{table}[!h]
\centering
\scalebox{1}{
\begin{tabular}{lcc}
\hline
                 & $l (^\circ )$ & $b (^\circ )$  \\
\hline
$H_0$-map$^{48}$ / $\gamma=65^{\circ}$ & $323.87 \pm 22.5$  & $28.04 \pm 22.5$  \\

$H_0$-map$^{192}$ / $\gamma=65^{\circ}$ & $328.25 \pm 11.2$ & $27.42 \pm 11.2$ \\

\hline
$H_0$-map$^{48}$ / $\gamma=70^{\circ}$ & $323.87 \pm 22.5$  & $28.04 \pm 22.5$  \\

$H_0$-map$^{192}$ / $\gamma=70^{\circ}$ & $326.63 \pm 11.2$ & $29.77 \pm 11.2$ \\
\hline
\end{tabular}
}
\caption{Robustness tests to verify our directional analysis results 
considering different spherical cap sizes.}
\label{results_table2}
\end{table}

\section{Consistency in the Calculation of the Effective Distance}\label{ap:eff_dist}

The calculation of the effective distance is an important step in our methodology of finding the velocity of the bulk flow. 
Equation~(\ref{Effectivedistance}) is based on the thermal noise model of~\cite{TURNBULL2012}, that needs the information of the peculiar velocities' uncertainties, which are individually unknown in the Pantheon+ catalog, and just appears as value $250$ ~km~s$^{-1}$ for all SNe. 

In Section~\ref{effdis-pecvel}, we have calculated $\sigma_{v_{pec}}$ using Equation~(\ref{eqsigma-v-pec}). 
However, one can adopt the Monte Carlo method to find $\sigma_{v_{pec}}$ and the effective distance $R$. 
For this, we simulate Gaussian distributions for each SN located within the spherical caps corresponding to both the $+/-$ dipole directions (see Table~\ref{results_table}), considering the value $\text{v}_{pec}^i$ (data in the Pantheon+ catalog) as the mean of the distribution and $250$ ~km~s$^{-1}$ as its standard deviation. 
For each distribution, one for each SN in the analysis, we randomly choose a value that we consider the uncertainty $\sigma_{v_{pec}}$ 
of such SN. 
The set of values $\{ \sigma_{v_{pec}} \}$ obtained in this form is termed one realization.  
With this set of uncertainties, we calculate the value of the effective distance $R$ using Equation~(\ref{Effectivedistance}) for each realization. 

\begin{figure}[!h]
\centering
\includegraphics[width=0.35\linewidth]{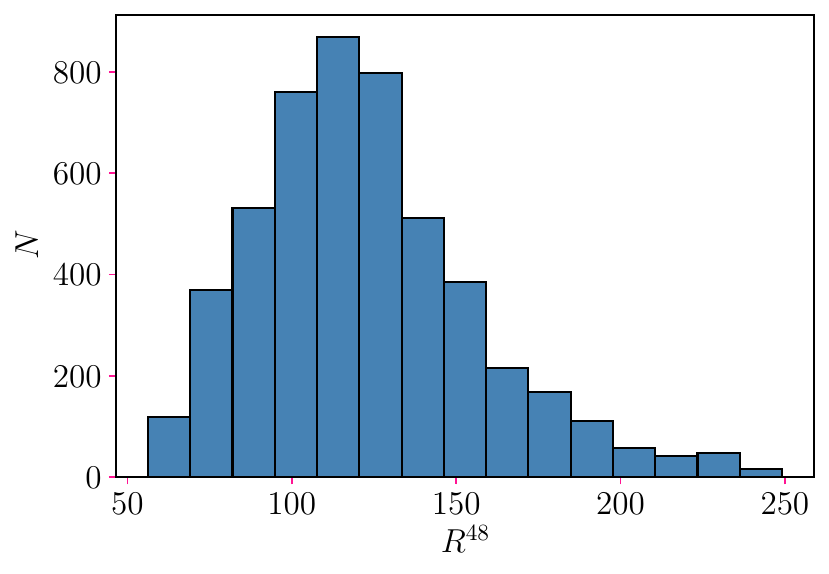}
\includegraphics[width=0.35\linewidth]{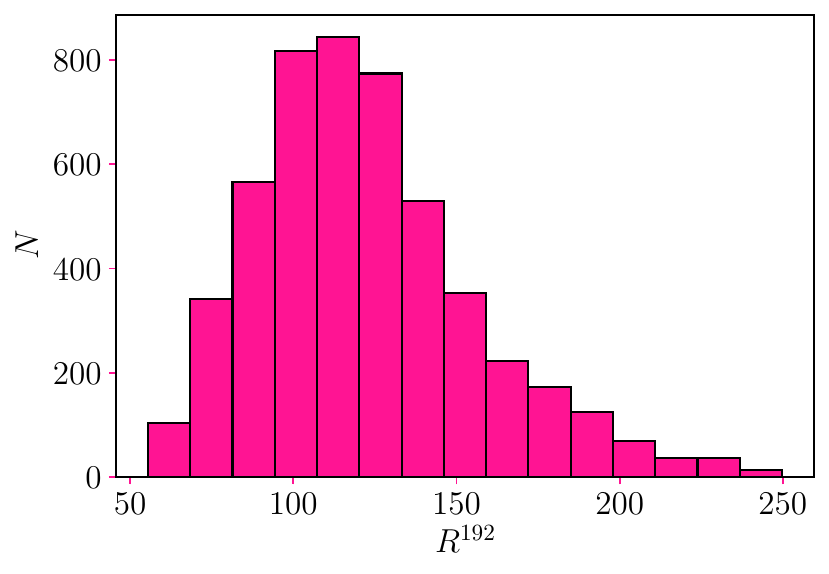}
\caption{Histograms of the distributions obtained using the Monte 
Carlo approach (see the text for details) to calculate the effective distance $R$, in units of Mpc, 
for $H_{0}$-map$^{48}$ (left  panel) and $H_{0}$-map$^{192}$ (right 
panel). 
The median values of these distributions, that provide our 
measurements of $R$, are: 
$R^{48} = 118.33 \pm 33.71$ Mpc, and 
$R^{192} = 117.57 \pm 33.60$ Mpc, respectively.
}
\label{histR}
\end{figure}

We repeat this procedure to produce a set of $N_{\scalebox{0.8}{realiz}} = 5000$ 
realizations\footnote{We verified that for a greater number of realizations the final value of the effective distance $R$ remains practically the same.}, for each angular resolution, that is, for $48$ and $192$ spherical 
caps. 
The set of effective distances obtained $\{ R_j \}, j=1,\cdots,5000$ 
produces two distributions for $48$ and $192$ caps, respectively, 
that we plot as histograms in Figure~\ref{histR}. 
The median of these distributions provides the value of the effective distance $R$. 
Therefore, we obtain $R^{48} = 118.33 \pm 33.71$ Mpc, and 
$R^{192} = 117.57 \pm 33.60$ Mpc,  
results that are in good agreement with those found in Section~\ref{effdis-pecvel}, with larger uncertainties than previous ones. 

\section{The Hubble Tension: Are Overdensities and Underdensities Biasing the \texorpdfstring{$H_0$}{} Measurements?}\label{ap:hubble-tension}

Recent works \citep{Giani24,Mazurenko24,Kenworthy19} have 
investigated the possibility that overdensities (i.e., large galaxy clusters) and underdensities (i.e., large voids) 
affect the $H_0$ measurements in the Local Universe, because such structures produce large inflows and outflows, respectively. 
In fact, this biasing effect exists, being important 
quantify the impact of such phenomena in the measurements of $H_0$. 
According to~\cite{Giani24}, the large supercluster Laniakea, which 
hosts the Milky Way, produces a negative average expansion of 
$\sim -1.1$ km~s$^{-1}$~Mpc$^{-1}$, inducing a variation in the Hubble constant $\Delta H_0\approx 0.5$ km~s$^{-1}$~Mpc$^{-1}$. 
Therefore, the inflows produced by Laniakea can increase the Hubble tension when applying this correction to the SNe Ia dataset. 
On the other hand,~\cite{Mazurenko24} noticed that a large supervoid could be generating outflows on scales $100 - 250$ Mpc~$h^{-1}$ inducing a change in the bulk flow velocities at such scales, then decreasing the tension due to the measurements reported by \cite{Watkins23} and \cite{Whitford23}; data that we commented at the end of Section~\ref{sec:final}. 
Lastly, \cite{Kenworthy19} affirm that overdensities and underdensities can change the Hubble constant in $2.2 \%$, not affecting the Hubble tension.

In this context, our directional analysis can didactically explain these results, and also provide a quantification of this biasing effect. 
The color of each pixel in our $H_0$-maps (see, e.g., the left column in Figure~\ref{fig:H0maps}) is the manifestation of the dominance of inflows or outflows, phenomena that increases or decreases the recession speed of matter structures with respect to the Hubble flow. 
In fact, the prevalence of these phenomena in the Local Universe is determined by the large-scale distribution of clustered matter and large voids, which can vary substantially along different directions
as recently noticed by \cite{Franco24}. 
However, this diversity is captured by our methodology, which provides, for the $i$th spherical cap in analysis, the best fit $H_0^i$ from the HL diagram of those SNe in the $i$th cap; 
since the $i$th cap has its vertex at the position of pixel $i$, 
it will have the color corresponding to the dominant effect in that direction: reddish (bluish) for values above (below) the mean, and greenish for values close to the mean. 
The relative differences of the pixel values of the $H_0$-map with 
respect to the $H_0$ value shown in Table~\ref{results_H0} are 
$\lesssim 3 \%$ in all cases, i.e., considering different angular 
resolutions and cap sizes, in perfect agreement with the outcomes 
of \cite{Kenworthy19}.  

\bibliographystyle{abbrvnat} 
\bibliography{manuscript} %

\end{document}